\documentclass[12pt]{article}
\usepackage{jheppub}

\usepackage{color}
\usepackage{graphicx}
\usepackage{float}
\usepackage{multirow, bigdelim}		    % For delim-command
\usepackage{mathtools}                            % For starred environment
\usepackage{caption}
\usepackage{subcaption}
\usepackage{hyperref}
\usepackage{enumitem}                            % customize list environments
\usepackage{amsmath}
\usepackage{amssymb}

\renewcommand{\thefigure}{\arabic{figure}}

\makeatletter                                              % custom labels  
\newcommand{\plabel}[2]{%
   \protected@write \@auxout {}{\string \newlabel {#1}{{#2}{\thepage}{#2}{#1}{}} }%
   \hypertarget{#1}{#2}
}

\def\td{{\widetilde \delta}}

\def\beq{\begin{equation}}
\def\eeq{\end{equation}}
\def\beeq{\begin{eqnarray}}
\def\eeeq{\end{eqnarray}}

\def\to{\rightarrow}

\def\nn{\nonumber}

\def\ID{1 \kern -.45 em 1}

\def\bea{\begin{eqnarray}}
\def\eea{\end{eqnarray}}
\def\Eq#1{Eq.~(\ref{#1})}

\begin{document}
%%%%%%%%%%%%%%%%%%%%%%%%%%%%%%%%%%%%%%%%%%%%%%%%%%%%%%%%%
\preprint{IFIC/15-69}

\title{Numerical implementation of the Loop-Tree Duality method}

\author[a]{Sebastian Buchta,}
\author[b]{Grigorios Chachamis,}
\author[c]{Petros Draggiotis} 
\author[a]{and Germ\'an Rodrigo}

\affiliation[a]{Instituto de F\'{\i}sica Corpuscular, Universitat de Val\`{e}ncia 
-- Consejo Superior de Investigaciones Cient\'{\i}ficas, 
Parc Cient\'{\i}fic, E-46980 Paterna, Valencia, Spain.}
\affiliation[b]{Instituto de F\'isica Te\'orica UAM/CSIC \& Universidad Aut\'onoma de
Madrid, C/ Nicol\'as Cabrera 15, E-28049 Madrid, Spain}
\affiliation[c]{Institute of Nuclear and Particle Physics, NCSR "Demokritos", 
Agia Paraskevi, 15310, Greece}

\emailAdd{sbuchta@ific.uv.es} 
\emailAdd{grigorios.chachamis@csic.es} 
\emailAdd{petros.draggiotis@gmail.com}
\emailAdd{german.rodrigo@csic.es}

\date{\today}

\abstract{
We present a first numerical implementation of the Loop-Tree Duality (LTD) method 
for the direct numerical computation of multi-leg one-loop Feynman integrals. 
We discuss in detail the singular structure of the dual integrands and define 
a suitable contour deformation in the loop three-momentum space 
to carry out the numerical integration. Then, we apply the LTD method to the 
computation of ultraviolet and infrared finite integrals, and present 
explicit results for scalar integrals with up to five external legs (pentagons) 
and tensor integrals with up to six legs (hexagons). The LTD method 
features an excellent performance independently of the number of external legs.}

\maketitle

%----------------------------------------------------------------------------------------
%	Introduction
%----------------------------------------------------------------------------------------

\section{Introduction}
\label{sec:intro}

The recent discovery of the Higgs boson at the LHC represents a great success of the Standard Model (SM) of
elementary particles. With the new run the primary goal is to study its properties in detail and to detect possible 
extentions to the SM. Precise theory predictions are needed to achieve this goal, which calls for  calculations
at the next-to-leading order (NLO) and beyond for multi-leg processes.

The development of automated NLO tools has seen great progress in recent years. 
Computing higher-order corrections in QFT, in particular in QCD and in the EW sector of the SM 
is highly challenging. The complexity increases as the
number of external particles gets bigger and the order of the perturbative expansion. 
The task is far from trivial and each step presents
its own difficulties: one needs first to generate the virtual and real scattering 
amplitudes, then carry out the integration over the loop momenta for the virtual 
contribution and finally perform the phase-space integration for both real and virtual 
corrections after taking proper care so that the infrared divergencies cancel.
In particular, infrared singularities of the virtual contribution can be subtracted 
by using appropriate semi-analytical terms and combine them with the ones stemming from 
the real corrections to produce finite results~\cite{Catani:1996jh}.
Purely numerical approaches on the integration of loop momenta have been discussed extensively 
in the literature~\cite{Soper:1998ye,Soper:1999xk,Soper:2001hu,Kramer:2002cd,Ferroglia:2002mz,Nagy:2003qn,Nagy:2006xy,Moretti:2008jj,Gong:2008ww,Kilian:2009wy,Becker:2010ng,Becker:2012aqa,Becker:2012nk}.
The generation of amplitudes and calculation of cross-sections at one loop has seen great progress 
in recent years and algorithmic calculations at NLO are now considered standardised, 
based on purely numerical~\cite{Bevilacqua:2011xh,Cascioli:2011va,Cullen:2014yla} 
and a mix of analytical and numerical approaches~\cite{Frixione:2008ym,Gleisberg:2008ta,Alwall:2014hca}. 
Substantial progress has also been made at higher orders~\cite{Passarino:2001wv,Anastasiou:2007qb,Becker:2012bi}.

The loop--tree duality (LTD) method~\cite{Catani:2008xa,Rodrigo:2008fp,Bierenbaum:2010cy,Bierenbaum:2012th,Bierenbaum:2013nja,Buchta:2014dfa,Buchta:2014fva,Buchta:phd,Buchta:2015vha,Hernandez-Pinto:2015ysa}
establishes that generic loop quantities (loop integrals and scattering amplitudes) 
in any relativistic, local and unitary field theory can be written 
as a sum of tree-level-like objects obtained after making all possible cuts to 
the internal lines of the corresponding Feynman diagrams, with one single cut per loop
and integrated over a measure that closely resembles the phase-space of the 
corresponding real corrections~\cite{Catani:2008xa,Rodrigo:2008fp}. 
This duality relation is realised by a modification of 
the customary $+i0$ prescription of the Feynman propagators
and encodes the causal structure of the scattering amplitudes
in the expected way. The analysis of the singular behaviour 
of one-loop integrals and scattering amplitudes in this framework 
at the integrand level in the loop momentum space shows that there is 
a partial cancellation of singularities among different dual 
contributions such that physical infrared and threshold 
singularities remain restricted to a compact region of the loop 
three-momentum~\cite{Bierenbaum:2013nja,Buchta:2014dfa,Buchta:2014fva}.
This feature opens up the intriguing possibility that virtual and real radiative corrections 
can be brought together under a common integral and be treated 
simultaneously with Monte Carlo techniques though a convenient mapping of
the external momenta entering the virtual and real scattering 
amplitudes~\cite{Hernandez-Pinto:2015ysa}.

In this work, we present a first numerical implementation of 
the LTD method and we apply it to the computation of multi-leg one-loop scalar and 
tensor integrals. The outline of the paper is as follows. 
In Section~\ref{sec:duality} we review the LTD method at one-loop
and discuss the singular behaviour of the dual integrand 
in the loop momentum space. 
In Section~\ref{sec:cd} we introduce the contour deformation 
in the loop three-momentum space which is used for the numerical loop integration. 
We present explicit numerical results for various external momenta configurations,
in Section~\ref{sec:rs} for scalar integrals up to pentagons and
in Section~\ref{sec:rt} for up to rank 3 tensor integrals with five and six external legs. 
Finally, our conclusions are presented in Section~\ref{sec:conclusions}.

%%%%%%%%%%%%%%%%%%%%%%%%%%%%%%%%%%%%%%%%%%%%%%%%%%%%%%%%%%%%%%%%%%%%%%%%%%%%
%%%%%%%%%%%%%%%%%%%%%%%%%%%%%%%%%%%%%%%%%%%%%%%%%%%%%%%%%%%%%%%%%%%%%%%%%%%%
%%%%%%%%%%%%%%%%%%%%%%%%%%%%%%%%%%%%%%%%%%%%%%%%%%%%%%%%%%%%%%%%%%%%%%%%%%%%

\section{Loop-tree duality at one-loop}
\label{sec:duality}

%============================================================
\begin{figure}[h]
\begin{center}
\includegraphics[scale=1.2]{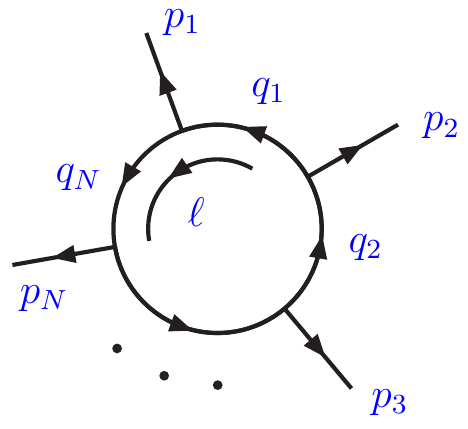}
\end{center}
\caption{\label{fig:1loop} 
{\em Momentum configuration of the one-loop $N$-point integral.}}
\end{figure}
%============================================================

We consider a general one-loop $N$-leg scalar integral (see Fig.~\ref{fig:1loop})
in dimensional regularization, with $d$  
the number of space-time dimensions,
\beq
\label{Ln}
L^{(1)}(p_1, p_2, \dots, p_N) =
\int_{\ell} \, \prod_{i \in \alpha_1} \, G_F(q_i)~, \qquad
\int_{\ell} \bullet =-i \int \frac{d^d \ell}{(2\pi)^{d}} \; \bullet~,
\eeq
where 
\beq
G_F(q_i)=\frac{1}{q_i^2-m_i^2+i0}
\label{eq:feynman}
\eeq
are Feynman propagators that depend on the 
loop momentum $\ell$, which flows anti-clockwise, 
and the four-momenta of the external legs $p_{i}$, 
$i \in \alpha_1 = \{1,2,\ldots N\}$, which are taken as outgoing and 
are clockwise ordered.
The momenta of the internal lines are denoted as $q_{i,\mu} = (q_{i,0},\mathbf{q}_i)$, 
where $q_{i,0}$ is the energy (time component) and $\mathbf{q}_{i}$ are 
the spatial components, which are defined as $q_{i} = \ell + k_i$ with 
$k_{i} = p_{1} + \ldots + p_{i}$, and $k_{N} = 0$ by momentum conservation. 
We also define $k_{ji} = q_j - q_i$ which, in fact, is independent of the loop 
momentum $\ell$.

The corresponding dual representation of the scalar integral in \Eq{Ln}
is obtained from the loop-tree duality (LTD) theorem~\cite{Catani:2008xa}: 
\bea
\label{oneloopduality}
L^{(1)}(p_1, p_2, \dots, p_N) 
&=& - \sum_{i \in \alpha_1} \, \int_{\ell} \; \td(q_i) \,
\prod_{\substack{j \in \alpha_1 \\ j\neq i}} \,G_D(q_i;q_j)~,
\eea 
where
\beq
G_D(q_i;q_j) = \frac{1}{q_j^2 -m_j^2 - i0 \,\eta k_{ji}}~,
\eeq
are dual propagators, $\eta$ is an arbitrary future-like vector, 
i.e.,~a $d$--dimensional vector that can be either light--like $(\eta^2=0)$ or
time--like $(\eta^2 > 0)$ with positive definite energy $\eta_0 \ge 0$, and 
\beq
\label{tdp}
\td(q_i) \equiv 2 \pi \, i \, \theta(q_{i,0}) \, \delta(q_i^2-m_i^2)~, 
\eeq
selects the internal loop on-shell modes, $G_F^{-1}(q_i)=0$, 
with positive definite energy, $q_{i,0}\geq 0$. Hence, the LTD theorem 
expresses the usual loop Feynman integral, \Eq{Ln},
as a sum of single-cut phase-space integrals, \Eq{oneloopduality}, 
with 
\beq
\int_{\ell} \td(q_i)~,
\eeq
as the single-particle phase-space integration measure. The LTD 
theorem is valid not only for scalar one-loop integrals, but
can straightforward be extended to deal with scattering 
amplitudes~\cite{Catani:2008xa} and higher orders of 
the perturbative expansion~\cite{Bierenbaum:2010cy,Bierenbaum:2012th}. 

The integrand of the dual representation of one-loop integrals 
or scattering amplitudes feature certain types of singularities 
leading to ultraviolet, infrared or threshold singularities.  
This singular behaviour has already been thoroughly discussed 
in~\cite{Buchta:2014dfa,Hernandez-Pinto:2015ysa}. 
We briefly recapitulate here the main points 
that are relevant in the present context. 

%============================================================
\begin{figure}[h]
\begin{center}
\includegraphics{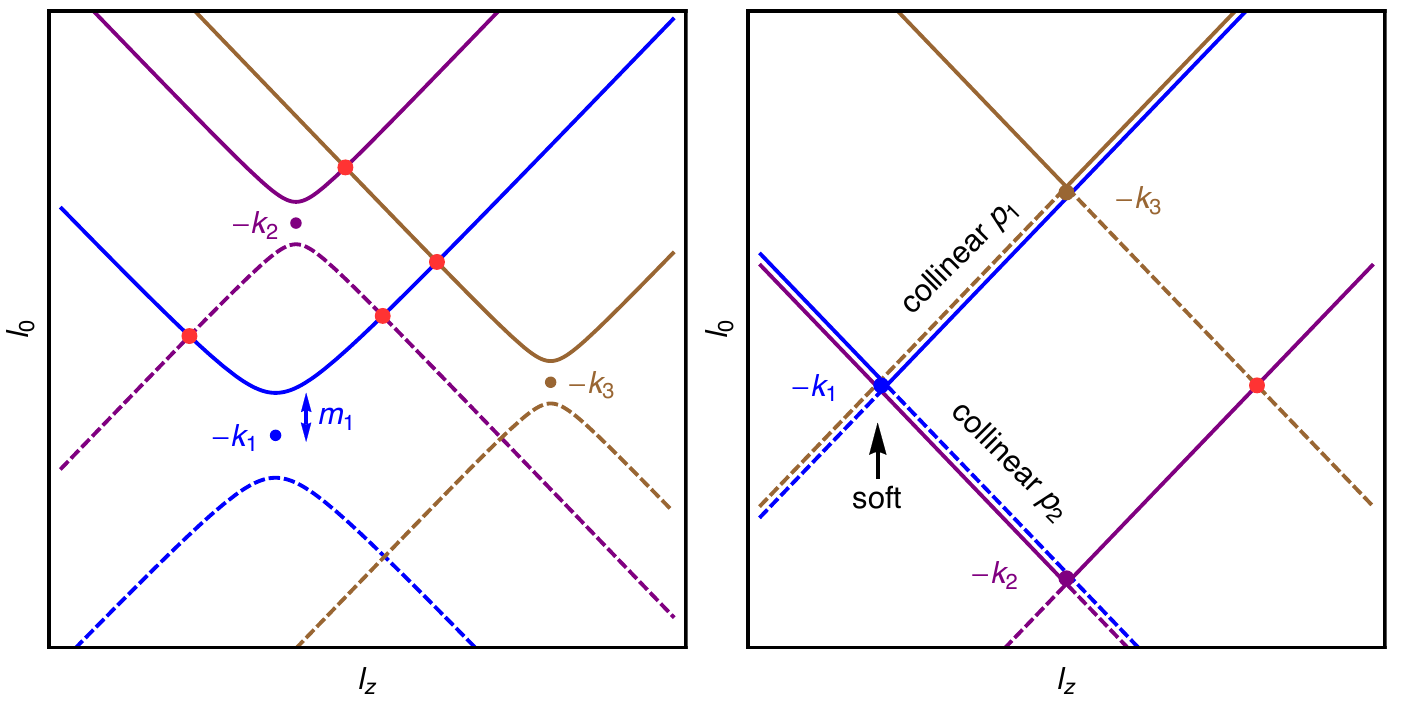}
\end{center}
\caption{On-shell hyperboloids for three arbitrary propagators 
in Cartesian coordinates in the ($\ell_0$,$\ell_z$) space (left). 
Kinematical configuration with massless propagators leading to infrared singularities (right).
In the latter case, the on-shell hyperboloids degenerate to light-cones. 
\label{fig:LCone}}
\end{figure}
%============================================================

For generic masses, the loop integrand in \Eq{Ln} becomes singular at the on-shell 
hyperboloids defined by $q_{i,0}^{(+)} = \sqrt{\mathbf{q}_i^2 + m_i^2-i0}$ 
(forward-hyperboloids, positive energy mode) and $q_{i,0}^{(-)} = -\sqrt{\mathbf{q}_i^2 + m_i^2-i0}$ 
(backward-hyperboloids, negative energy mode). This is illustrated in Fig.~\ref{fig:LCone}
for a given kinematical configuration with three internal loop 
propagators. Solid lines in Fig.~\ref{fig:LCone} represent the  
forward on-shell hyperboloids, and dashed lines the backward on-shell hyperboloids. 
The LTD method is equivalent to evaluating the sum of the integrals along the forward on-shell hyperboloids with the singularities appearing at the intersection of each forward on-shell 
hyperboloid with the forward of backward on-shell hyperboloid of the other propagators. 
A crucial point of this discussion is the observation that 
dual propagators can be rewritten as 
\beq
\td(q_i) \, G_D(q_i;q_j) = 
i\, 2 \pi \, 
\frac{\delta(q_{i,0}-q_{i,0}^{(+)})}{2 q_{i,0}^{(+)}} \, 
\frac{1}{(q_{i,0}^{(+)} + k_{ji,0})^2-(q_{j,0}^{(+)})^2}~,
\label{eq:newdual}
\eeq
where $q_{i,0}^{(+)}$ can be interpreted as the loop energy measured along 
the forward on-shell hyperboloid with origin at $-k_i$. 
From \Eq{eq:newdual} it is obvious that dual propagators become 
singular, $G_D^{-1}(q_i;q_j)=0$, if one of the following conditions is fulfilled:
\bea
& q_{i,0}^{(+)}+q_{j,0}^{(+)}+k_{ji,0}=0~, \label{ellipsoid} \\
& q_{i,0}^{(+)}-q_{j,0}^{(+)}+k_{ji,0}=0~. \label{hyperboloid}
\eea
The first condition, \Eq{ellipsoid}, is satisfied if the 
forward on-shell hyperboloid of $G_F(q_i)$ intersects 
with the backward on-shell hyperboloid of $G_F(q_j)$. 
The second condition, \Eq{hyperboloid}, is true when 
the two forward on-shell hyperboloids intersect each other. 

The solution to \Eq{ellipsoid} is an ellipsoid in the loop
three-momentum space and requires $k_{ji,0}<0$.
Moreover, since it is the result of the intersection of a forward 
with a backward on-shell hyperboloid the distance between the two 
propagators has to be future-like, $k_{ji}^2 \ge 0$. 
Actually, internal masses restrict this condition to
\beq
k_{ji}^2-(m_j+m_i)^2 \ge 0~, \qquad k_{ji,0}<0~, \qquad 
\rm{forward~with~backward~hyperboloids}~.
\label{eq:generalizedtimelike}
\eeq
The second equation, \Eq{hyperboloid}, leads to a hyperboloid
in the loop three-momentum space, and there are solutions for $k_{ji,0}$ 
either positive or negative, namely when either of the two momenta is 
set on-shell. 
Here, the distance between the momenta of the propagators 
has to be space-like, although also time-like configurations 
can fulfill~\Eq{hyperboloid} as far as the time-like 
distance is small or close to light-like: 
\beq
k_{ji}^2-(m_j-m_i)^2 \le 0~, \qquad \rm{two~forward~hyperboloids}~.
\label{eq:generalizedspacelike}
\eeq

As it was demonstrated in~\cite{Buchta:2014dfa}, the 
integrand singularities appearing from the intersection of forward with 
forward on-shell hyperboloids cancel among dual contributions. 
To see that one needs to keep in mind that  propagators are positive inside 
the on-shell hyperboloids and negative outside. When integrating along 
the forward on-shell hyperboloids, every singularity is crossed twice. Firstly 
when going from the inside to the outside (or from the outside to the inside) 
and secondly from the outside to the inside (or from the inside to the outside). 
The crucial point is that the 
contributions coming from the two integrands have opposite sign and thus 
cancel out. Note that the imaginary dual prescription $\eta\cdot k_{ji}$ changes 
sign from the one dual contribution to the other to ensure the cancellation
of the singularities. On the contrary, the singularities from the intersection 
of a forward with a backward on-shell hyperboloid survive because only 
a single dual contribution leads to that singularity and there is no possibility 
of cancellation. In the case of integrable singularities, a contour deformation
can be employed as explained in the next section.

\begin{figure}[htb]
\begin{center}
\begin{subfigure}[c]{.2\textwidth}
   \bea
   \left( \begin{matrix}
   0 & \text{\bf{H}} &  \text{\bf{E}} \\
    \text{\bf{H}} & 0 &  \text{\bf{E}} \\
   0 & 0 & 0
   \end{matrix} \right) ~~= \nn
   \eea
\end{subfigure}
\begin{subfigure}[c]{.75\textwidth}
   \includegraphics[width=\linewidth]{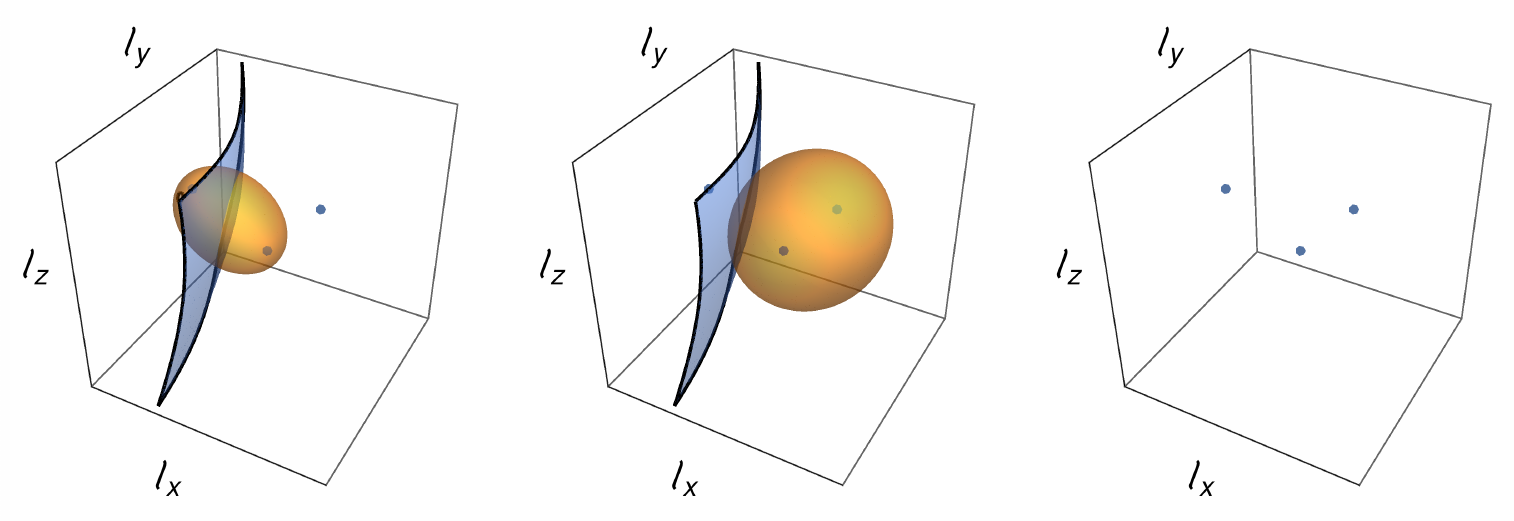}
\end{subfigure}
\caption{Singularity matrix of a sample three-point function
and the corresponding singularity surfaces of the dual integrands
in the loop three-momentum space. 
\label{fig:trilms}}
\end{center}
\end{figure}

\begin{figure}[htb]
\begin{center}
\begin{subfigure}[c]{.24\textwidth}
   \bea
   \left( 
   \begin{matrix}
   0 & \text{\bf{H}} &  \text{\bf{H}} &  \text{\bf{E}} \\
    \text{\bf{H}} & 0 &  \text{\bf{H}} & \text{\bf{E}} \\
    \text{\bf{H}} & \text{\bf{H}} & 0 & \text{\bf{E}} \\
   0 & 0 & 0 & 0
   \end{matrix} \right) ~~= \nn
   \eea
\end{subfigure}
\begin{subfigure}[c]{.75\textwidth}
   \includegraphics[width=\linewidth]{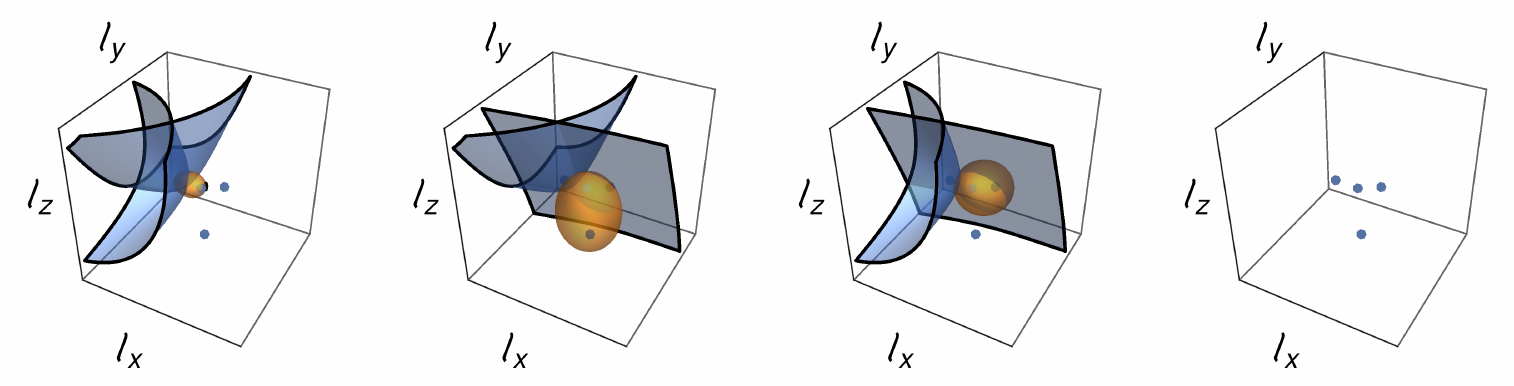}
\end{subfigure}
\caption{Singularity matrix of a sample four-point function
and the corresponding singularity surfaces of the dual integrands
in the loop three-momentum space.
\label{fig:boxlms}}
\end{center}
\end{figure}

The action of the LTD can be encoded symbolically by the following matrix scheme
\bea
G_F\cdot G_F \cdots G_F\quad\xrightarrow{\hspace*{1.0cm}\text{\normalsize LTD}\hspace*{1.0cm}} \qquad
\left(\begin{matrix} 
\delta & G_D & G_D & \cdots & G_D\\
 G_D & \delta & G_D & \cdots & G_D\\
 G_D & G_D & \delta & \cdots & G_D\\
 \vdots & \vdots & \vdots & \ddots & \vdots\\
 G_D & G_D & G_D & \cdots & \delta
 \end{matrix} \right)
 \label{fig:ltdsymb}
  \eea
Each line in the matrix to the right of the arrow in~\Eq{fig:ltdsymb} 
represents a dual contribution with one single propagator on-shell, 
$\delta=\td(q_i)$. The column index points to the 
corresponding dual propagators, $G_D=G_D(q_i;q_j)$. 
This scheme can now be used to graphically indicate the position of  different 
singularities in a given dual integral. In Figs.~\ref{fig:trilms} and~\ref{fig:boxlms}
we apply it to a triangle and a box respectively. To be more specific, in Fig.~\ref{fig:trilms} 
each of the 3D plots in the r.h.s represents the singularities of any
one dual contribution. We plot the ellipsoid (orange surfaces) 
and hyperboloid (blue surfaces) singularities in the loop three-momentum space. 
The blue dots are the foci of the on-shell hyperboloids, i.e. 
$-\mathbf{k}_i,\,\,\, i\in \{1,2,3\}$.  In the l.h.s, we see the singularity scheme, 
where the first line of the matrix corresponds to the first plot in the r.h.s, the second
line corresponds to the second 3D plot and so on.
In the matrix, an H 
indicates that the corresponding dual propagator from~\Eq{fig:ltdsymb} 
generates an hyperboloid singularity, E stands for ellipsoid singularities, 
and zero means no singularity. 
Similarly, for a four-point function in Fig. \ref{fig:boxlms}.

In both cases, the hyperboloid singularities always appear pairwise across the dual 
contributions. This is not by accident.  Due to the symmetry of \Eq{hyperboloid}
under the exchange of $i$ ($i$ counts dual contributions) and 
$j$ ($j$ counts leg positions) the hyperboloid singularities 
always appear in pairs and are distributed symmetrically around 
the main diagonal. Inspecting \Eq{ellipsoid}, which is the defining equation for 
ellipsoid singularities we see that this equation is not 
symmetric under the exchange of indices. Thus for every 
ellipsoid singularity we have a  zero as its counterpart.

%%%%%%%%%%%%%%%%%%%%%%%%%%%%%%%%%%%%%%%%%%%%%%%%%%%%%%%%%
%%%%%%%%%%%%%%%%%%%%%%%%%%%%%%%%%%%%%%%%%%%%%%%%%%%%%%%%%
%%%%%%%%%%%%%%%%%%%%%%%%%%%%%%%%%%%%%%%%%%%%%%%%%%%%%%%%%
%%%%%%%%%%%%%%%%%%%%%%%%%%%%%%%%%%%%%%%%%%%%%%%%%%%%%%%%%

%%%%%%%%%%%%%%%%%%%%%%%%%%%%%%%%%%%%%%%%%%%%%%%%%%%%%%%%%
%%%%%%%%%%%%%%%%%%%%%%%%%%%%%%%%%%%%%%%%%%%%%%%%%%%%%%%%%
%%%%%%%%%%%%%%%%%%%%%%%%%%%%%%%%%%%%%%%%%%%%%%%%%%%%%%%%%
%%%%%%%%%%%%%%%%%%%%%%%%%%%%%%%%%%%%%%%%%%%%%%%%%%%%%%%%%

At this point,
we have established that 
the hyperboloid  singularities do cancel among dual contributions and therefore we do 
not need to treat them in any special manner. Still though, they do have an impact on the 
way we need to deform our contour. 
This is due to the fact that in order to preserve the cancellation of 
hyperboloid singularities, dual contributions featuring the same 
hyperboloid singularity must receive the same deformation. To 
further illustrate this point, let us look at the pentagon example 
shown in Fig.~\ref{fig:pentacouple}. In Fig. \ref{fig:pentacouple}, contributions one, 
two and three are coupled via their common hyperboloid singularities. 
Thus, they need to receive the very same deformation that accounts 
for all the ellipsoid singularities occurring within those contributions. 
These are found at position four of the second contribution 
and positions one and four of the third contribution. The fourth 
dual contribution is not coupled to any other contribution and a standalone
deformation can be applied. The fifth contribution does not require any 
treatment.

%%%%%%%%%%%%%%%%%%%%%%%%%%%%%%%%%%%%%%%%%%%%%%%%%%%%%%
%%%%%%%%%%%%%%%%%%%%%%%%%%%%%%%%%%%%%%%%%%%%%%%%%%%%%%
\begin{figure}[htb]
\bea
\left( 
\begin{matrix*}[l]
0 & \text{\bf{H}} & 0 & 0 & 0 \\
\text{\bf{H}} & 0 & \text{\bf{H}} & \text{\bf{E}} & 0 \\
\text{\bf{E}} & \text{\bf{H}} & 0 & \text{\bf{E}} & 0 \\
\text{\bf{E}} & \text{\bf{E}} & \text{\bf{E}} & 0 & \text{\bf{E}} \\
0 & 0 & 0 & 0 & 0 
\end{matrix*}
\right)
\begin{matrix*}[l]
& \rdelim\}{3}{10mm}[] & \text{Contributions are coupled:}\\
& & \text{Every contribution receives all deformations}\\
& & \text{that occur within the group.}\\
&  \; \to\; & \text{Deform with ellipsoids that itself contains.}\\
&  \;\to\; & \text{No deformation needed here.}
\end{matrix*}\nn
\eea
\caption{Five-point function with dual contributions coupled by hyperboloid singularities.
\label{fig:pentacouple}}
\end{figure}
%%%%%%%%%%%%%%%%%%%%%%%%%%%%%%%%%%%%%%%%%%%%%%%%%%%%%%
%%%%%%%%%%%%%%%%%%%%%%%%%%%%%%%%%%%%%%%%%%%%%%%%%%%%%%

As a general strategy, one organises the dual contributions into groups. 
A group is a set of pairwise coupled contributions. To each of the groups a contour 
deformation is applied independently from the others. Within a group every contribution 
receives the same deformation that accounts for all the ellipsoids of the group.
Turning back to the example in Fig.~\ref{fig:pentacouple}, we have three groups: the first 
group involves contributions one to three, the second group is 
contribution four and the third group is contribution five.

%----------------------------------------------------------------------------------------
%	Contour deformation
%----------------------------------------------------------------------------------------

\section{ The deformation of the contour }
\label{sec:cd}

\begin{figure}[htb]
\centering
\includegraphics[scale=0.75]{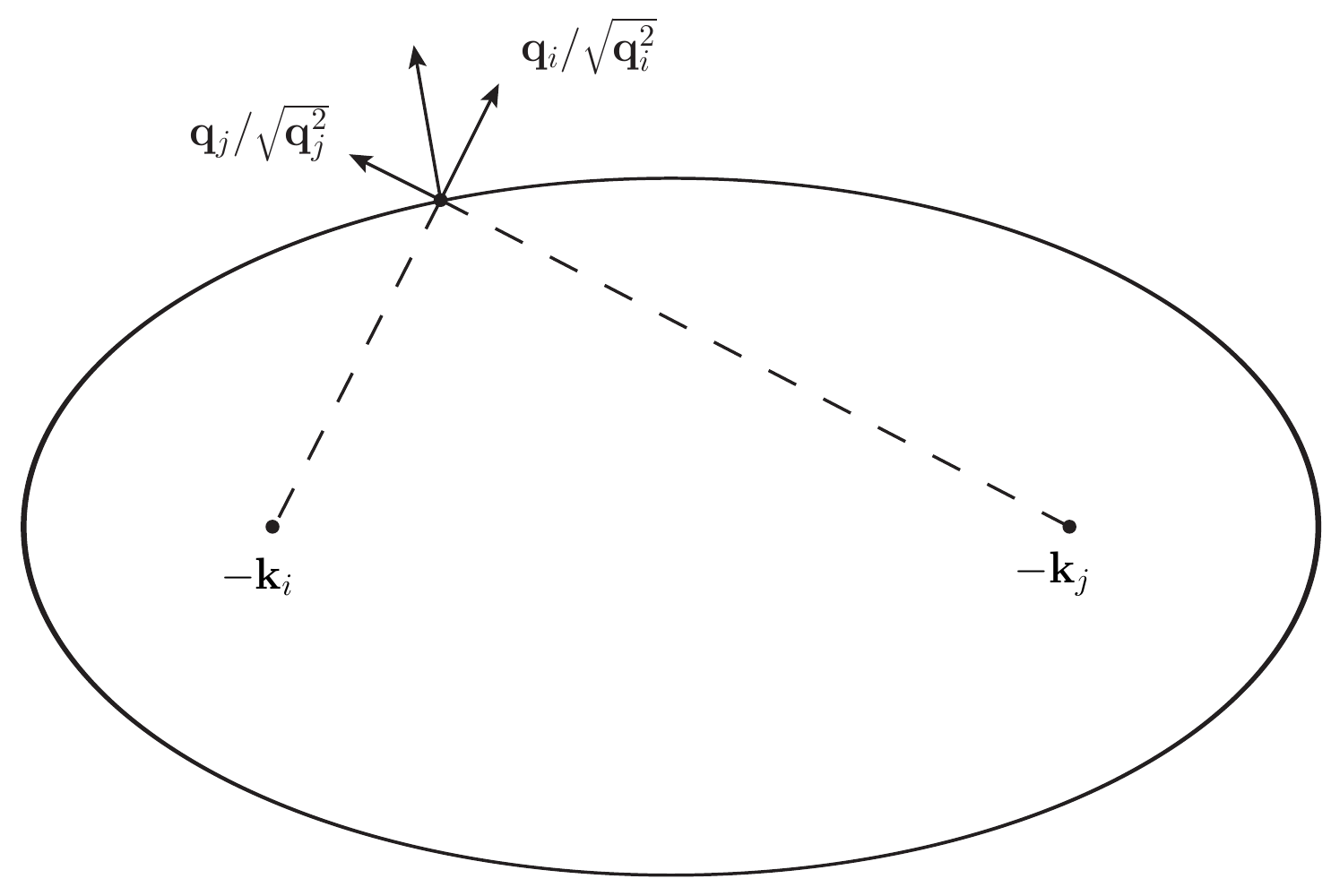}
\caption{Two-dimensional slice of the ellipsoid singularity  
of dual contribution $i$ at position $j$. The resulting vector 
gives the direction of the deformation.
\label{fig:vectorpart}}
\end{figure}

As we saw in Section \ref{sec:duality}, the ellipsoid singularities (forward--backward type) lead to integrable threshold singularities 
that lie on the real axis. To deal with them, we need to deform the integration path into 
the imaginary space. Every valid deformation must satisfy a certain set of 
requirements \cite{Soper:1999xk}:
\begin{enumerate}
\item The deformation has to respect the $i0$-prescription of the propagator.
In general, a contour deformation in the loop three-momentum space has the form:
\bea
\boldsymbol{\ell}\to\boldsymbol{\ell}'=\boldsymbol{\ell}+i\boldsymbol{\kappa}~.
\label{eq:gendef}
\eea
where $\boldsymbol{\kappa}$ is a function of the loop momentum $\ell$ and the external momenta. 
In our case, we want to perform the integration over a product of dual propagators. 
Inserting~\Eq{eq:gendef} into the on-shell energy relation, we obtain
\bea
q_{i,0}^{(+)}=\sqrt{-\boldsymbol{\kappa}^2+
2i\boldsymbol{\kappa}\cdot\mathbf{q}_i+\mathbf{q}_i^2+m_i^2-i0}~.
\eea
The $i0$-prescription tells us in which direction to deform when coming 
close to a singularity. Hence, any valid deformation must match this prescription. 
Consequently we need to have
\bea
\boldsymbol{\kappa}\cdot\mathbf{q}_i<0~.
\label{eq:cond1}
\eea
\item The deformation should vanish at infinity:
We are looking for a deformation that does not change the actual value of the integral. 
We do not want $|\kappa|$ to grow for $|\ell|\to\infty$. An easy 
way to satisfy this condition is to choose $\kappa$ such that $|\kappa|\to 0$ 
as $|\ell |\to \infty$.\footnote{
Strictly speaking, there is a third condition.
The deformation must vanish at the position of soft or collinear singularities.
This point is of importance for the matching of soft and collinear singularities 
between real and virtual corrections~\cite{Hernandez-Pinto:2015ysa}. 
If the deformation shifts those singularities 
together with everything else, the cancellation will be spoiled.
However, in the scope of this article, we are only dealing with infrared 
finite diagrams.}
\end{enumerate}
With these conditions in mind, we construct the deformation in the following way:
As explained in Section \ref{sec:duality}, we first organise the dual contributions 
into groups. For every ellipsoid singularity of the group we include a factor:
\bea
\lambda_{ij}\left(\frac{\mathbf{q}_i}{\sqrt{\mathbf{q}_i^2}}+
\frac{\mathbf{q}_j}{\sqrt{\mathbf{q}_j^2}}\right)\exp\left(
- \frac{G_D^{-2}(q_i;q_j)}{A_{ij}}\right)~,
\label{eq:deffac}
\eea
with $\mathbf{q}_i=\boldsymbol{\ell}+\mathbf{k}_i$ and $\boldsymbol{\ell}$ 
the loop three-momentum. The deformation factor in \Eq{eq:deffac} consists of 
two main components. 
The first component defines the direction of the deformation, and is given by the sum 
of the two unit vectors $\mathbf{q}_i/\sqrt{\mathbf{q}_i^2}$ and 
$\mathbf{q}_j/\sqrt{\mathbf{q}_j^2}$. As shown in Fig.~\ref{fig:vectorpart}
the vectors $\mathbf{q}_i$ and $\mathbf{q}_j$ have their origin in $-\mathbf{k}_i$ 
and $-\mathbf{k}_j$, respectively, and the deformation 
is designed to point to the outside of the singularity ellipsoid.
For an efficient numerical implementation, however, we should also take into account 
in the selection of the deformation parameters that for massive propagators the vectors
$-\mathbf{k}_i$ and $-\mathbf{k}_j$ might be slightly displaced from the true 
focal points of the ellipsoid.
Inside the  ellipsoid, the sum of the two unit vectors 
$\mathbf{q}_i/\sqrt{\mathbf{q}_i^2}$ and $\mathbf{q}_j/\sqrt{\mathbf{q}_j^2}$
helps to flatten the deformation and indeed they cancel each other along 
the major axis of the ellipsoid. 
By choosing all the scaling parameters $\lambda_{ij}<0$ for all possible 
combinations $\{ij\}$ we satisfy the first condition in \Eq{eq:cond1}. 

%%%%%%%%%%%%%%%%%%%%%%%%%%%%%%%%%%%%%%%%%%%%%%%%%%%%%%%%%
%%%%%%%%%%%%%%%%%%%%%%%%%%%%%%%%%%%%%%%%%%%%%%%%%%%%%%%%%
%%%%%%%%%%%%%%%%%%%%%%%%%%%%%%%%%%%%%%%%%%%%%%%%%%%%%%%%%
%%%%%%%%%%%%%%%%%%%%%%%%%%%%%%%%%%%%%%%%%%%%%%%%%%%%%%%%%

The second component in \Eq{eq:deffac} is the exponential factor $\exp(-G_D^{-2}(q_i;q_j)/A_{ij})$ 
that suppresses the deformation at infinity. At singular points, $G_D^{-2}(q_j;q_i)$ 
vanishes and thus the deformation reaches its maximum. Far away form 
the singularity, $-G_D^{-2}(q_j;q_i)$ reaches a large negative value and thus the 
exponential tends rapidly to zero. Finally, the factor $\lambda_{ij}$ is a scaling factor,
and $A_{ij}$ determines the width of the deformation. 
The indices $ij$ in $\lambda_{ij}$ and $A_{ij}$ indicate that those 
parameters can be selected individually for each deformation contribution
for optimization purposes. Then, we sum over the entire group of 
coupled singularities and arrive at:
\bea
\boldsymbol{\kappa} = \sum\limits_{i,j\in \text{group}}
\lambda_{ij}\left(\frac{\mathbf{q}_i}{\sqrt{\mathbf{q}_i^2}}+
\frac{\mathbf{q}_j}{\sqrt{\mathbf{q}_j^2}}\right)\exp\left(
- \frac{G_D^{-2}(q_j;q_i)}{A_{ij}}\right)~.
\label{eq:defogroup}
\eea
The corresponding Jacobian can be calculated analytically or numerically; in our 
current implementation we have chosen the analytic way.

%----------------------------------------------------------------------------------------
%	Results for scalar one-loop, multi-leg integrals
%----------------------------------------------------------------------------------------

\section{Results for multi-leg scalar one-loop integrals}
\label{sec:rs}

We have implemented the LTD method in a 
\verb!C++!  code and all the results in this paper were
obtained on a desktop machine with an 
Intel i7 (3.4GHz) processor with 8 cores and 16 GB of RAM.
The program uses the Cuba library~\cite{Hahn:2004fe} 
as a numerical integrator. 
The user needs only to input the number of 
external legs, the external momenta, the internal masses and has
the freedom to change the parameters of the contour deformation. The momenta and 
masses can be read in from a text file. 
The user can choose between Cuhre~\cite{Cuhre1, Cuhre2} and VEGAS~\cite{Lepage:1980dq}, 
and give the desired number of evaluations. 
At run time, the code performs the following steps:
\begin{enumerate}
\item Reads in masses and external momenta.
\item Checks where ellipsoid singularities occur.
\item Checks where hyperboloid singularities occur, groups the dual contributions 
accordingly and applies the contour deformation.
\item Calls the integrator.
\end{enumerate}
We use {\tt MATHEMATICA 10.0}~\cite{Mathematica} 
to generate random momenta and masses 
to scan as much of the phase-space as possible, to ensure that 
the program works properly in all regions. 
For our numerical results, the routine Cuhre was used unless otherwise stated.
The momenta 
and masses of all the sample phase-space points used in the 
following sections are collected in Appendix~\ref{AppendixA}.
We mainly used {\tt LoopTools 2.10}~\cite{Hahn:1998yk}
and also {\tt SecDec 3.0}~\cite{Borowka:2015mxa} 
to produce reference results to compare with.

%-----------------------------------
%	Scalar Triangles
%-----------------------------------

\subsection{Scalar triangles}

We consider first infrared finite scalar triangle integrals. 
The sample point P\ref{point1} in Table~\ref{tab:tridef} has all internal masses equal 
while P\ref{point2} has three different internal masses. Momenta 
and masses were chosen randomly between $-100$~GeV and $+100$~GeV. 
Similarly, P\ref{point3} in Table \ref{tab:tridef} 
has all internal masses equal whereas in P\ref{point4} all three of them 
have different values. 

Points with momentum configurations that do not need deformation 
(i.e. whose loop integral is purely real) are computed in well below one second 
with a precision of at least 4 digits. 
For points with momentum configurations that require deformation,
the calculation time increases to typically $3-15$ seconds.

%%%%%%%%%%%%%%%%%%%%%%%%%%%%%%%%%%%%%%%%%%%%%%%%%%%%
%%%%%%%%%%%%%%%%%%%%%%%%%%%%%%%%%%%%%%%%%%%%%%%%%%%%
\begin{table}[htb]
\begin{center}
\begin{tabular}{|llll|} \hline
& Scalar Triangle & Real Part  &  Imaginary Part \\
\hline
P\plabel{point1}{1} 
   & LoopTools & $-5.85694 \times 10^{-5}$ & \\
   & LTD       & $-5.85685(24) \times 10^{-5}$ & \\
\hline
P\plabel{point2}{2} 
   & LoopTools & $-3.39656 \times 10^{-7}$ & \\
   & LTD       & $-3.39688(53) \times 10^{-7}$ & \\
\hline
P\plabel{point3}{3} 
   & LoopTools & $~~5.37305 \times 10^{-4}$    & $- i~6.68103 \times 10^{-4}$ \\
   & LTD       & $~~5.37307(9) \times 10^{-4}$ & $- i~6.68103(9) \times 10^{-4}$ \\
\hline
P\plabel{point4}{4} 
   & LoopTools & $-5.61370 \times 10^{-7}$ & $- i~1.01665 \times 10^{-6}$ \\
   & LTD       & $-5.61363(83) \times 10^{-7}$ & $- i~1.01666(8) \times 10^{-6}$ \\
\hline
\end{tabular}
\caption{Sample scalar triangles.
\label{tab:tridef}}
\end{center}
\end{table}
%%%%%%%%%%%%%%%%%%%%%%%%%%%%%%%%%%%%%%%%%%%%%%%%%%%%
%%%%%%%%%%%%%%%%%%%%%%%%%%%%%%%%%%%%%%%%%%%%%%%%%%%%

\begin{figure}[htb]
\begin{center}
   \includegraphics[width=0.48\linewidth]{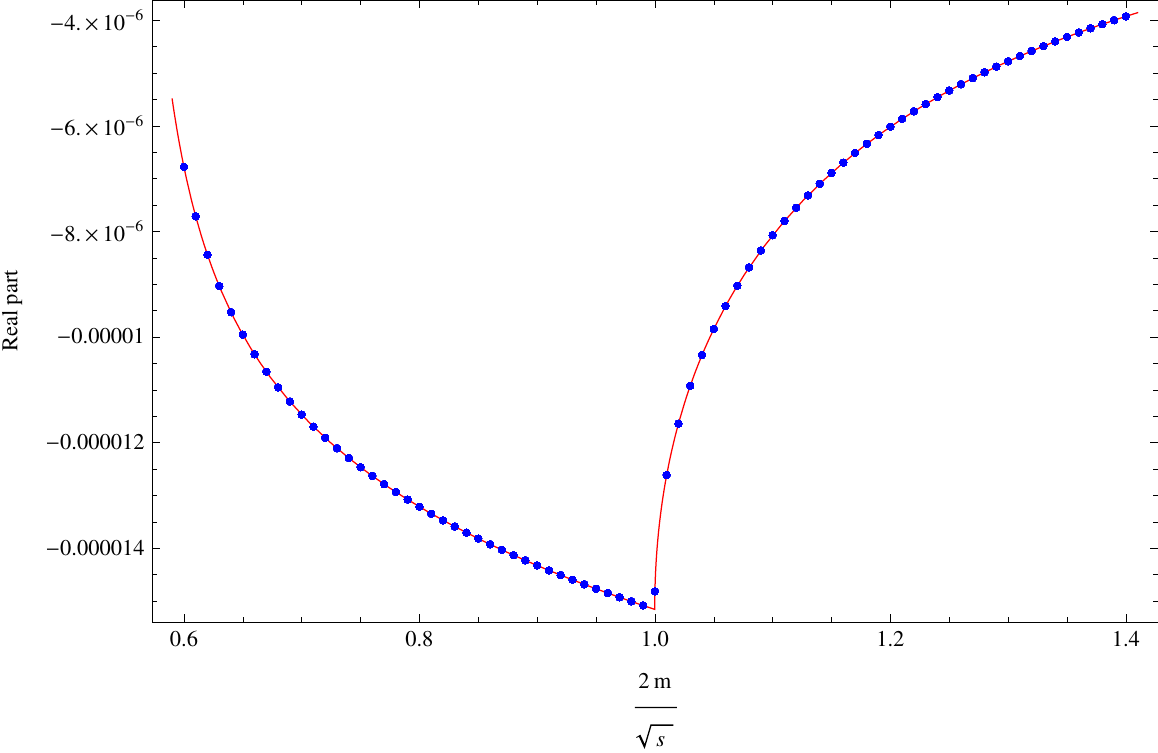}
   \includegraphics[width=0.48\linewidth]{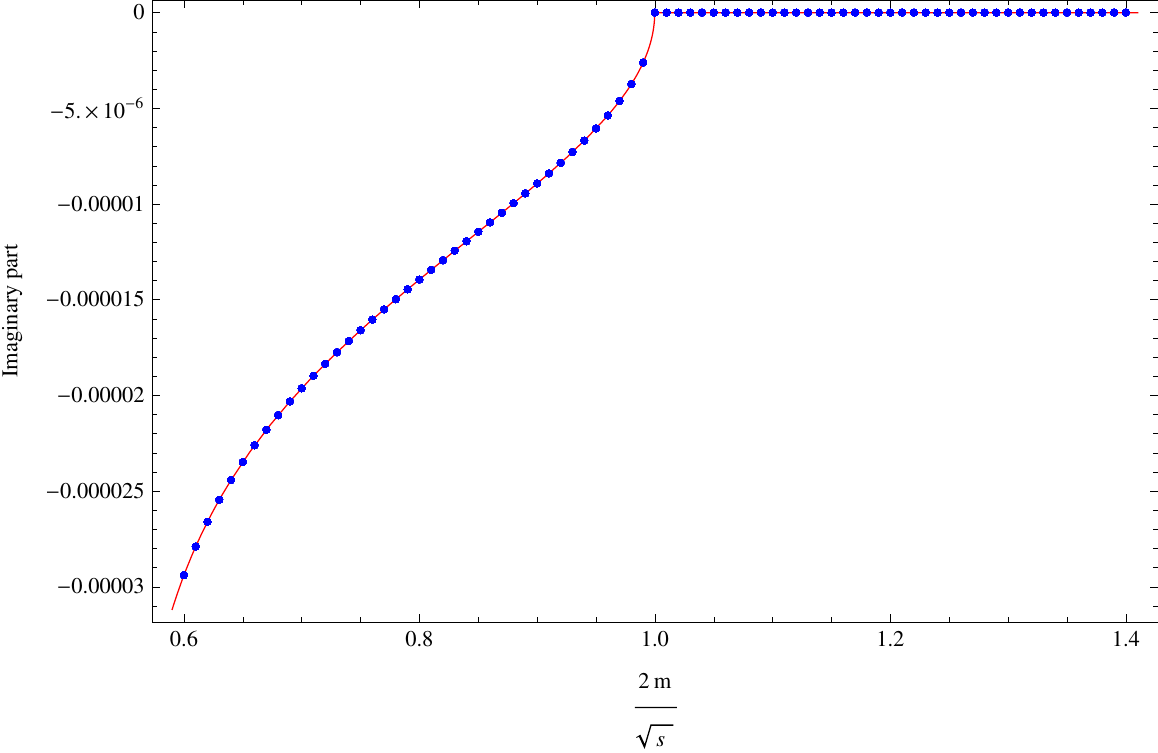}
\caption{Mass scan of the region around threshold. The red curve 
is done with LoopTools and the blue points are obtained with the LTD method.
\label{fig:trithreshold}}
\end{center}
\end{figure}

An important check of our implementation
is the mass-scan around threshold. 
In Fig.~\ref{fig:trithreshold} all internal masses are equal, i.e.
$m_i = m$, $i\in\{1,2,3\}$, and the center-of-mass energy $s$ was kept 
constant while varying the mass $m$. 
The calculation time remains 
constant for all mass values.

%%%%%%%%%%%%%%%%%%%%%%%%%%%%%%%%%%%%%%%%%%%%%%%%%%%%
%%%%%%%%%%%%%%%%%%%%%%%%%%%%%%%%%%%%%%%%%%%%%%%%%%%%
%%%%%%%%%%%%%%%%%%%%%%%%%%%%%%%%%%%%%%%%%%%%%%%%%%%%
%%%%%%%%%%%%%%%%%%%%%%%%%%%%%%%%%%%%%%%%%%%%%%%%%%%%

%-----------------------------------
%	Scalar Boxes
%-----------------------------------

\subsection{Scalar Boxes}

Next we consider infrared finite box scalar integrals. 
To get good precision (4 digits), for  boxes that need deformation we use 
$4$ to $5\cdot 10^6$ Cuhre calls, whereas for phase-space 
points with no deformation only $5\cdot 10^4$ calls, the same as in the triangle case. 
This is reflected in the running times. 
Points with deformation require  about 15 seconds whereas points with
no deformation well below one second.
While it is practically guaranteed to get the no-deformation points 
with good precision, 
for points with deformation the quality of results depends on the proper choice of 
the deformation parameters. Therefore, we mainly focus our 
attention to such points in the following.
The sample points P\ref{point5} and P\ref{point7} of Table \ref{tab:boxes} 
correspond to a momentum configuration in which all four internal masses are equal. In 
P\ref{point6} and P\ref{point8} all masses are different. In P\ref{point9} two 
adjacent internal lines have equal masses as well as the two opposing 
ones. P\ref{point10} represents a situation in which opposite 
lines have equal masses. 

%%%%%%%%%%%%%%%%%%%%%%%%%%%%%%%%%%%%%%%%%%%%%%%%%%%%
%%%%%%%%%%%%%%%%%%%%%%%%%%%%%%%%%%%%%%%%%%%%%%%%%%%%
\begin{table}[htb]
\begin{center}
\begin{tabular}{|llll|} \hline
&  & Real Part  &  Imaginary Part \\ 
\hline
P\plabel{point5}{5} & LoopTools & 2.15339$\times 10^{-13}$ & \\
                    & LTD & 2.15319(52)$\times 10^{-13}$ & \\
\hline
P\plabel{point6}{6} & LoopTools & 1.39199$\times 10^{-11}$ & \\
                    & LTD & 1.39199(6)$\times 10^{-11}$ & \\
\hline
P\plabel{point7}{7} & LoopTools & -2.38766$\times 10^{-10}$ & -3.03080$\times 10^{-10}$\\
                    & LTD & -2.38775(76)$\times 10^{-10}$ & -3.03063(76)$\times 10^{-10}$\\
\hline
P\plabel{point8}{8} & LoopTools & -4.27118$\times 10^{-11}$ & 4.49304$\times 10^{-11}$\\
                    & LTD & -4.27120(95)$\times 10^{-11}$ & 4.49307(95)$\times 10^{-11}$\\
\hline
P\plabel{point9}{9} & LoopTools & -7.37897$\times 10^{-11}$  & -1.19657$\times 10^{-10}$\\
                    & LTD & -7.37916(782)$\times 10^{-11}$ & -1.19649(78)$\times 10^{-10}$\\
\hline
P\plabel{point10}{10} & LoopTools & -1.85544$\times 10^{-10}$ & 2.13553$\times 10^{-10}$\\
                    & LTD & -1.85548(8)$\times 10^{-10}$ & 2.13554(8)$\times 10^{-10}$\\
\hline
\end{tabular}
\caption{Sample scalar boxes.
\label{tab:boxes}}
\end{center}
\end{table}
%%%%%%%%%%%%%%%%%%%%%%%%%%%%%%%%%%%%%%%%%%%%%%%%%%%%
%%%%%%%%%%%%%%%%%%%%%%%%%%%%%%%%%%%%%%%%%%%%%%%%%%%%

\begin{figure}[htb]
\begin{center}
   \includegraphics[width=0.48\linewidth]{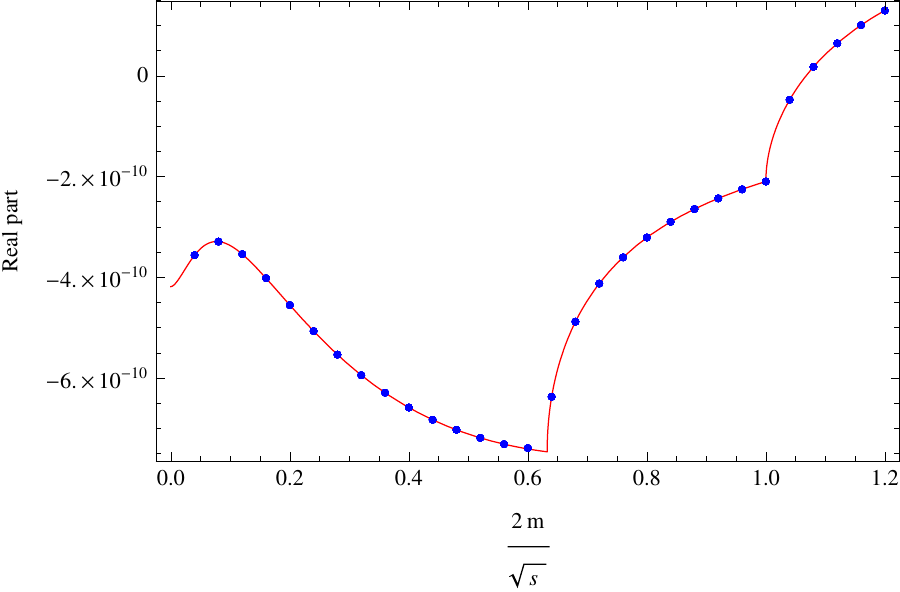}
   \includegraphics[width=0.48\linewidth]{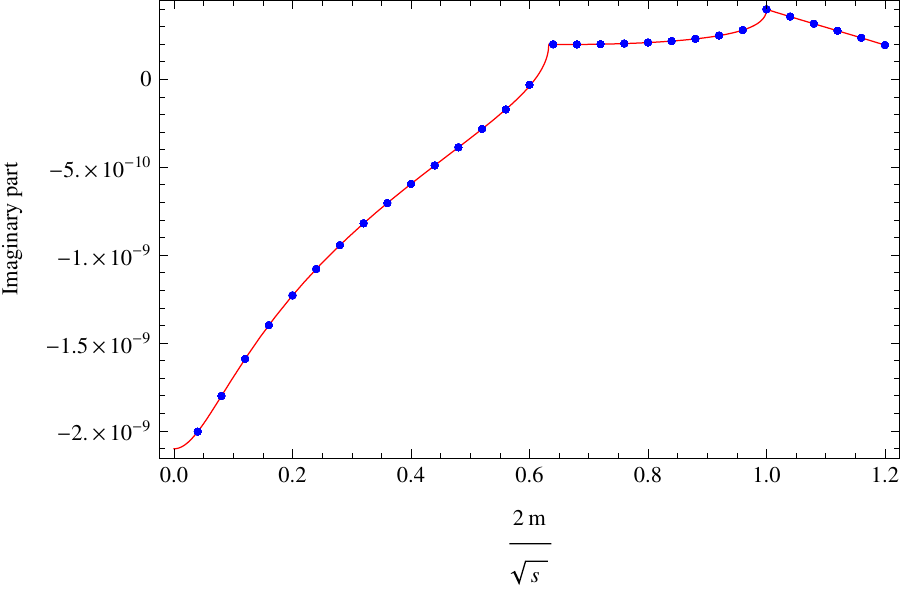}
\caption{Mass-scan of a box integral. The red curve 
is done with LoopTools and the blue points are obtained with the LTD method. 
\label{fig:boxalleq}}
\end{center}
\end{figure}

We perform again a mass-scan (see Fig. \ref{fig:boxalleq}) 
with all internal masses  equal, i.e. $m_i = m, i\in{1,2,3,4}$. 
The center-of-mass energy $s$ was kept constant while the mass $m$ was varied. 
The program deals well
with all kinds of boxes, even when many different kinematical scales are involved.
In Fig. \ref{fig:boxalleq}, two thresholds are crossed at $2m/\sqrt{s}=0.65$ 
and $1$. From right to left, the number of ellipsoid singularities grows by 
one after each threshold is crossed, starting from one to end up to three.

%%%%%%%%%%%%%%%%%%%%%%%%%%%%%%%%%%%%%%%%%%%%%%%%%%%%
%%%%%%%%%%%%%%%%%%%%%%%%%%%%%%%%%%%%%%%%%%%%%%%%%%%%
%%%%%%%%%%%%%%%%%%%%%%%%%%%%%%%%%%%%%%%%%%%%%%%%%%%%
%%%%%%%%%%%%%%%%%%%%%%%%%%%%%%%%%%%%%%%%%%%%%%%%%%%%

%-----------------------------------
%	Scalar Pentagons
%-----------------------------------

\subsection{Scalar Pentagons}

Let us now turn to pentagon diagrams. 
No-deformation points are computed with $10^5$ evaluations 
which takes about $0.5$ seconds. Points with deformation 
demand $5\cdot 10^6$ evaluations to maintain the level of precision 
of the triangles and boxes. This results to an average calculation 
time of about 30 seconds. 

%%%%%%%%%%%%%%%%%%%%%%%%%%%%%%%%%%%%%%%%%%%%%%%%%%%%
%%%%%%%%%%%%%%%%%%%%%%%%%%%%%%%%%%%%%%%%%%%%%%%%%%%%
\begin{table}[htb]
\begin{center}
\begin{tabular}{|llll|} \hline
& Scalar Pentagon & Real Part  &  Imaginary Part \\
\hline
P\plabel{point11}{11} 
   & LoopTools & $-1.24025 \times 10^{-13}$ & \\
   & LTD       & $-1.24027(16) \times 10^{-13}$ & \\
\hline
P\plabel{point12}{12} 
   & LoopTools & $-1.48356 \times 10^{-14}$ & \\
   & LTD       & $-1.48345(116) \times 10^{-14}$ & \\
\hline
P\plabel{point13}{13} 
   & LoopTools & $~~1.02350 \times 10^{-11}$    & $+ i~1.40382 \times 10^{-11}$ \\  
   & LTD       & $~~1.02353(1) \times 10^{-11}$ & $+ i~1..40385(1) \times 10^{-11}$ \\
\hline
P\plabel{point14}{14} 
   & LoopTools & $-1.52129 \times 10^{-15}$      & $- i~1.17401 \times 10^{-14}$ \\
   & LTD       & $-1.52657(602) \times 10^{-15}$ & $- i~1.17483(60) \times 10^{-15}$ \\
\hline
P\plabel{point15}{15} 
   & LoopTools & $-4.29464 \times 10^{-15}$    & $- i~6.55440 \times 10^{-14}$ \\
   & LTD       & $-4.29520(845) \times 10^{-15}$ & $- i~6.55433(85) \times 10^{-14}$ \\
\hline
\end{tabular}
\caption{Sample scalar pentagons.
\label{tab:pentex}}
\end{center}
\end{table}
%%%%%%%%%%%%%%%%%%%%%%%%%%%%%%%%%%%%%%%%%%%%%%%%%%%%
%%%%%%%%%%%%%%%%%%%%%%%%%%%%%%%%%%%%%%%%%%%%%%%%%%%%

\begin{figure}[htb]
\begin{center}
   \includegraphics[width=0.48\linewidth]{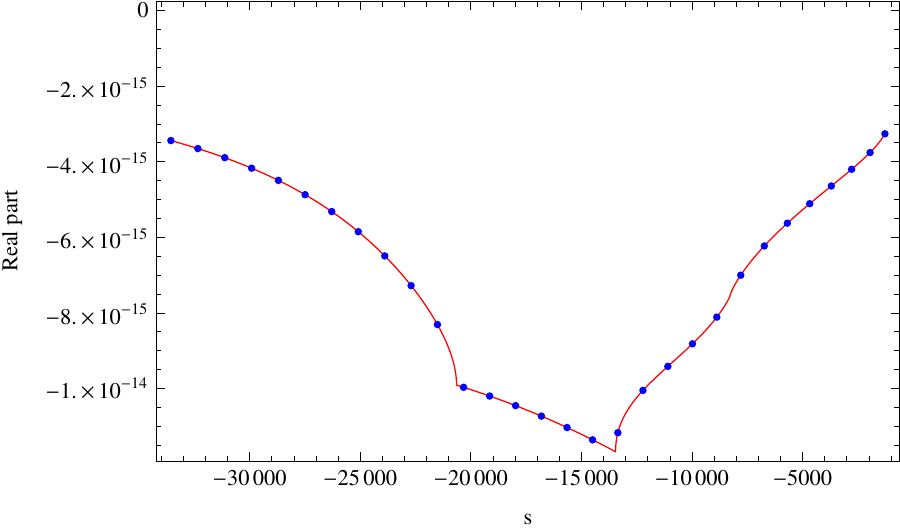}
   \includegraphics[width=0.48\linewidth]{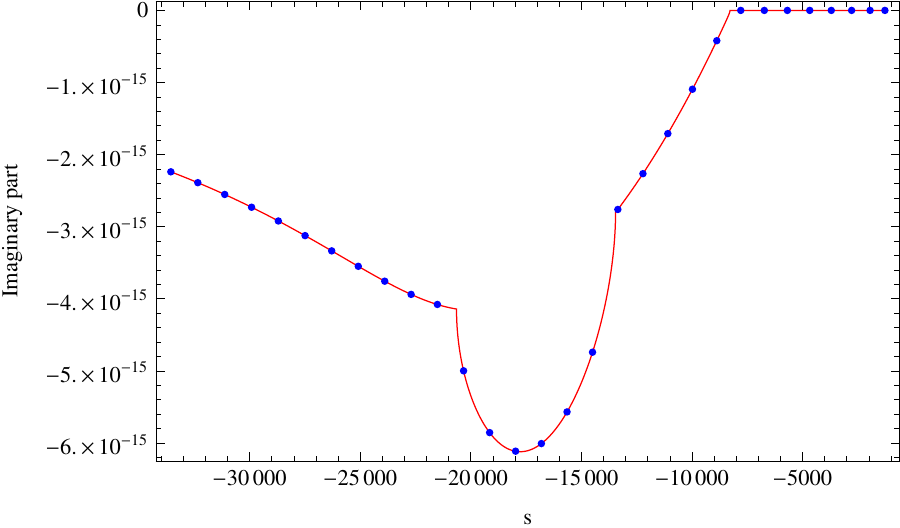}
\caption{Energy-scan of a scalar pentagon. The red curve 
is done with LoopTools and the blue points are obtained with the LTD method.
\label{fig:pentachu}}
\end{center}
\end{figure}

In Table~\ref{tab:pentex} we display a collection of pentagon example 
results for different kinematical configurations. In P\ref{point11} and 
P\ref{point13} all internal masses are equal; in P\ref{point14} they are all distinct from 
each other and in P\ref{point15} we have $m_1=m_2=m_3\neq m_4=m_5$. 
Our implementation of the LTD method shows its robustness by producing accurate 
results regardless of the kinematical situation. This statement is further 
supported by an energy-scan which we performed and which is shown in Fig.~\ref{fig:pentachu}.
The center-of-mass energy $s$ is varied. This is realized by 
varying $p_3$ while keeping $p_3^2$ constant. Of course, due to 
momentum conservation, this involves 
$p_4^2=(p_1+p_2+p_3)^2$ not being constant. 
In this scan, we cross three thresholds at $s\approx -8.5\cdot 10^3, -13.5\cdot 
10^3$ and $-21\cdot 10^3$ GeV$^2$ which divide the scan into four zones. From 
right to left, we start with zero ellipsoid singularities in the first zone, then we have 
one in the second zone, two in the third zone and finally one in the leftmost zone.

%%%%%%%%%%%%%%%%%%%%%%%%%%%%%%%%%%%%%%%%%%%%%%%%%%%%
%%%%%%%%%%%%%%%%%%%%%%%%%%%%%%%%%%%%%%%%%%%%%%%%%%%%
%%%%%%%%%%%%%%%%%%%%%%%%%%%%%%%%%%%%%%%%%%%%%%%%%%%%
%%%%%%%%%%%%%%%%%%%%%%%%%%%%%%%%%%%%%%%%%%%%%%%%%%%%

%----------------------------------------------------------------------------------------
%	Results for tensor integrals
%----------------------------------------------------------------------------------------

\section{Tensor loop integrals}
\label{sec:rt}

The LTD relation for scalar loop integrals can easily be extended to deal 
tensor integrals. As long as the quantum 
field theory is local and unitary, these tensor factors do not lead to 
additional singularities~\cite{Catani:2008xa} and 
the LTD method can then be applied in a straightforward manner.  
If the one-loop integral features a non-trivial numerator $\mathcal{N}(\ell,\{p_i\})$, 
\bea
L^{(1)}(p_1, \dots, p_N; \mathcal{N}(\ell,\{p_i\})) 
= \int_{\ell} \, \mathcal{N}(\ell,\{p_i\}) \,
\prod_{i\in\alpha_1} \,G_F(q_i)~,
\eea 
then, the LTD theorem takes the form
\bea
L^{(1)}(p_1, \dots, p_N; \mathcal{N}(\ell,\{p_i\})) 
= - \sum_{i\in\alpha_1} \, \int_{\ell} \, \td(q_i) \,
\mathcal{N}(\ell,\{p_i\}) \,
\prod_{\substack{j\in\alpha_1 \\ j\neq i}} \,G_D(q_i;q_j)~.
\eea 
While the numerator is formally left unchanged, there is actually a 
potential implication. The presence of the dual delta function 
demands $q_{i,0}^{(+)}=\sqrt{\mathbf{q}_i^2+m_i^2-i0}$, which is equivalent to 
\bea
\ell_0=-k_{i,0}+\sqrt{\mathbf{q}_i^2+m_i^2-i0}~.
\label{eq:l0num}
\eea
In other words, whenever we perform a single cut of a Feynman graph, 
the numerator has to be evaluated at the position of the cut which is 
fixed by the dual delta function. As a direct consequence, the numerator 
takes a different form in each dual contribution.

Another important aspect to take into consideration is the cancellation of 
singularities among dual contributions.  Here, we would like to make 
explicit that the numerators do not spoil 
the cancellation of the hyperboloid singularities.
A typical numerator is a polynomial of 
scalar products of the loop-momentum with external momenta: $\ell\cdot p_k$. 
Let us see what happens to a single factor 
when it hits the singularity. Note first, that the hyperboloid singularity is 
given by \Eq{hyperboloid} which we rewrite in the more suitable form
\bea
q_{i,0}^{(+)}-k_{i,0} = q_{j,0}^{(+)}-k_{j,0}~.
\label{eq:hyperboloidnewform}
\eea
Using \Eq{eq:l0num}, the loop-momentum $\ell$ contracted with some 
external momentum $p_k$ is:
\bea
\ell\cdot p_k\,|_{\text{i-th cut}} = (q_{i,0}^{(+)}-k_i) p_{k,0}-\boldsymbol{\ell}\cdot \mathbf{p}_{k}
= (q_{j,0}^{(+)}-k_j) p_{k,0}-\boldsymbol{\ell}\cdot \mathbf{p}_{k} = \ell\cdot p_k\,|_{\text{j-th cut}}~,
\label{eq:cancelnum}
\eea
where we have used \Eq{eq:hyperboloidnewform}. 
This means that the numerators of two 
dual contributions $i$ and $j$ take the same value at their common pole, 
thus leaving the cancellation of hyperboloid singularities intact. 
This is an important property to take advantage of, because 
it allows us to straightforwardly apply the LTD method to such diagrams 
without any additional effort.

%----------------------------------------------------------------------------------------
%	Tensor Pentagons
%----------------------------------------------------------------------------------------

\subsection{Tensor Pentagons}

Next, we investigate tensor pentagon  integrals at the one-loop level with 
numerators up to rank three. 
The number of evaluations is chosen to be the same as in the 
scalar case, i.e. $10^5$ times for no-deformation points and $5\cdot 10^6$  times for 
phase-space points that require deformation. This results in calculation times 
of about 1 second and 30 seconds, respectively. 

In Table~\ref{tab:pentagonlp} we show a selection of sample points.
The reference points P\ref{point24} and P\ref{point26} feature the rank two numerator 
$(\ell\cdot p_3)\times(\ell\cdot p_4)$ while P\ref{point25} 
and P\ref{point27} have the numerator $(\ell\cdot p_3)\times(\ell\cdot p_4)\times(\ell\cdot p_5)$. 
All the points have all internal masses equal.  P\ref{point27} 
actually contains six ellipsoid singularities 
whereas the other sample points have two to three.
We include this point to demonstrate that 
the program does well even under such challenging circumstances.

For tensor pentagons and hexagons, we have used the program SecDec~\cite{Borowka:2015mxa} 
to cross-check our results. We have run SecDec taking no care to optimise
its runtime. This means that in the following, whenever we present the
running times of SecDec we do it for completeness reasons and not because
we imply that our code compares better or worse to SecDec. A proper
comparison of our implementation with available codes is beyond the scope
of this paper.

%%%%%%%%%%%%%%%%%%%%%%%%%%%%%%%%%%%%%%%%%%%%%%%%%%%%
%%%%%%%%%%%%%%%%%%%%%%%%%%%%%%%%%%%%%%%%%%%%%%%%%%%%
\begin{table}[htb]
\begin{center}
\begin{tabular}{|ccllll|}
\hline
 & Rank & Tensor Pentagon & Real Part  &  Imaginary Part & Time [s]\\
\hline
P\plabel{point24}{16} & 2 & LoopTools & $-1.86472\times 10^{-8}$ & & \\
 & & SecDec & $-1.86471(2)\times 10^{-8}$ & & 45\\
 & & LTD       & $-1.86462(26)\times 10^{-8}$ & & 1\\
\hline
P\plabel{point25}{17} & 3 & LoopTools & $~~1.74828\times 10^{-3}$ & & \\
 & & SecDec & $~~1.74828(17)\times 10^{-3}$ & & 550\\
 & & LTD       & $~~1.74808(283)\times 10^{-3}$ & & 1\\
\hline
P\plabel{point26}{18} & 2 & LoopTools & $-1.68298\times 10^{-6}$ & $+i~1.98303\times 10^{-6}$ &\\
 & & SecDec & $-1.68307(56)\times 10^{-6}$ & $+i~1.98279(90)\times 10^{-6}$ & 66\\
 & & LTD       & $-1.68298(74)\times 10^{-6}$ & $+i~1.98299(74)\times 10^{-6}$ & 36\\
\hline
P\plabel{point27}{19} & 3 & LoopTools & $-8.34718\times 10^{-2}$ & $+i~1.10217\times 10^{-2}$ & \\
 & & SecDec & $-8.33284(829)\times 10^{-2}$ & $+i~1.10232(107)\times 10^{-2}$ & 1501\\
 & & LTD       & $-8.34829(757)\times 10^{-2}$ & $+i~1.10119(757)\times 10^{-2}$ & 38\\
\hline
\end{tabular}
\caption{Tensor pentagons involving numerators of rank two and three.}
\label{tab:pentagonlp}
\end{center}
\end{table}
%%%%%%%%%%%%%%%%%%%%%%%%%%%%%%%%%%%%%%%%%%%%%%%%%%%%
%%%%%%%%%%%%%%%%%%%%%%%%%%%%%%%%%%%%%%%%%%%%%%%%%%%%

\begin{figure}[htb]
\begin{center}
   \includegraphics[width=0.48\linewidth]{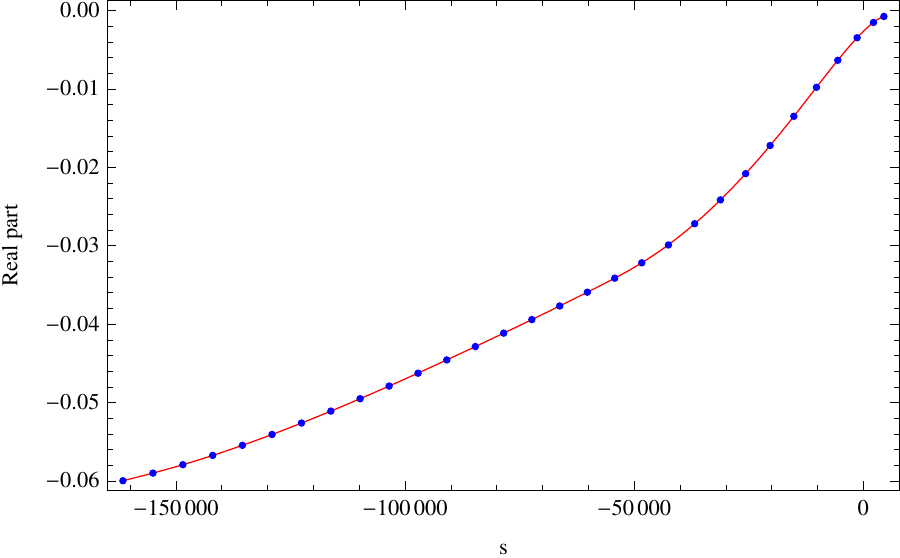}
   \includegraphics[width=0.48\linewidth]{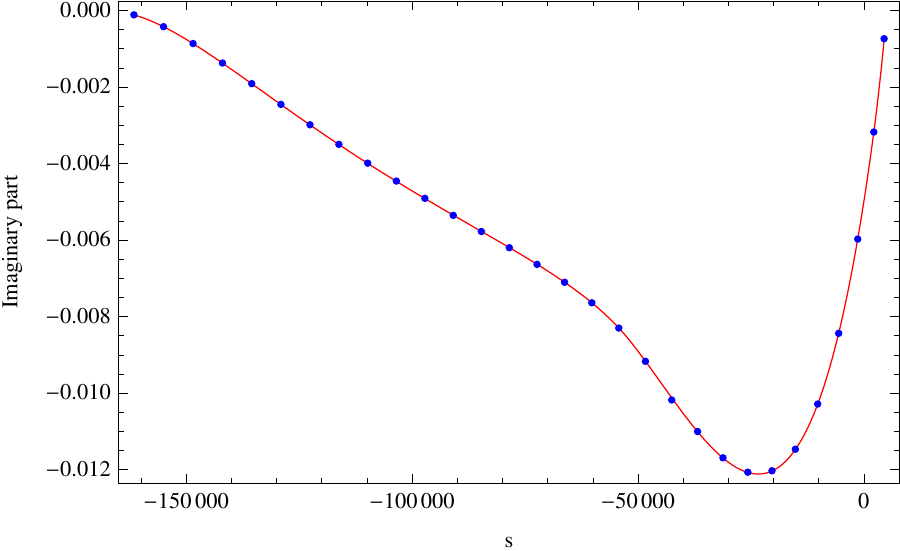}
\caption{Energy-scan of a rank three tensor pentagon around the threshold region. 
The red curve is done with LoopTools and the blue points are obtained with the LTD method.
\label{fig:pentanum}}
\end{center}
\end{figure}

We have performed several different scans; a sample is presented in Fig.~\ref{fig:pentanum}. 
In this energy-scan, we varied $p_1$ and thus the center-of-mass energy $s=(p_1+p_2)^2$, 
similar to what we have done with scalar pentagons. The corresponding numerator 
function is $(\ell\cdot p_1)\times(\ell\cdot p_2)\times(\ell\cdot p_3)$, 
which means that both numerator and denominator change in the scan. 
In Fig.~\ref{fig:pentanum}, one can see that the LTD method is able to 
successfully pass this  test.

%%%%%%%%%%%%%%%%%%%%%%%%%%%%%%%%%%%%%%%%%%%%%%%%%%%%
%%%%%%%%%%%%%%%%%%%%%%%%%%%%%%%%%%%%%%%%%%%%%%%%%%%%
%%%%%%%%%%%%%%%%%%%%%%%%%%%%%%%%%%%%%%%%%%%%%%%%%%%%
%%%%%%%%%%%%%%%%%%%%%%%%%%%%%%%%%%%%%%%%%%%%%%%%%%%%

%----------------------------------------------------------------------------------------
%	Tensor Hexagons
%----------------------------------------------------------------------------------------

\subsection{Tensor Hexagons}

Finally, we  compute hexagon tensor integrals. 
The number of evaluations for no-deformation points is $10^6$ and 
for deformation points $8\cdot 10^6$. 
The typical corresponding computation times are about 10 and 75 seconds, respectively. 
Since LoopTools can provide results only up to pentagons, we 
used exclusively here the program SecDec to cross-check.

%%%%%%%%%%%%%%%%%%%%%%%%%%%%%%%%%%%%%%%%%%%%%%%%%%%%
%%%%%%%%%%%%%%%%%%%%%%%%%%%%%%%%%%%%%%%%%%%%%%%%%%%%
\begin{table}[htb]
\begin{center}
\begin{tabular}{|ccllll|} \hline
& Rank & Tensor Hexagon & Real Part  &  Imaginary Part & Time[s]\\
\hline
P\plabel{point28}{20} 
   & 1 & SecDec    & $-1.21585(12) \times 10^{-15}$ & & 36\\
   &   & LTD       & $-1.21552(354)\times 10^{-15}$ & & 6\\
\hline
P\plabel{point29}{21} 
   & 3 & SecDec    & $~~4.46117(37) \times 10^{-9}$ & & 5498\\
   &   & LTD       & $~~4.461369(3) \times 10^{-9}$ & & 11\\
\hline
P\plabel{point30}{22} 
   & 1 & SecDec    & $~~1.01359(23) \times 10^{-15}$ & $+ i~2.68657(26) \times 10^{-15}$ & 33\\
   &   & LTD       & $~~1.01345(130)\times 10^{-15}$ & $+ i~2.68633(130)\times 10^{-15}$ & 72\\
\hline
P\plabel{point31}{23} 
   & 2 & SecDec    & $~~2.45315(24) \times 10^{-12}$ & $- i~2.06087(20) \times 10^{-12}$ & 337\\
   &   & LTD       & $~~2.45273(727)\times 10^{-12}$ & $- i~2.06202(727) \times 10^{-12}$ & 75\\
\hline
P\plabel{point32}{24} 
   & 3 & SecDec    & $-2.07531(19) \times 10^{-6}$ & $+ i~6.97158(56) \times 10^{-7}$ & 14280\\
   &   & LTD       & $-2.07526(8)  \times 10^{-6}$ & $+ i~6.97192(8) \times 10^{-7}$ & 85\\
\hline
\end{tabular}
\caption{Tensor hexagons involving numerators of rank one to three.
\label{tab:hexagonlp}}
\end{center}
\end{table}
%%%%%%%%%%%%%%%%%%%%%%%%%%%%%%%%%%%%%%%%%%%%%%%%%%%%
%%%%%%%%%%%%%%%%%%%%%%%%%%%%%%%%%%%%%%%%%%%%%%%%%%%%

We present a selection of sample points in Table \ref{tab:hexagonlp}. 
P\ref{point28} and P\ref{point30} feature the rank one numerator 
$\ell\cdot p_1$, the former has all internal masses different, in the latter 
they are all equal. P\ref{point29} has six distinct internal masses and 
the numerator function  $(\ell\cdot p_2)\times(\ell\cdot p_4)\times(\ell\cdot p_6)$,
P\ref{point31} possesses the numerator $(\ell\cdot p_2)\times(\ell\cdot p_5)$ and six 
different masses, as well. Finally, P\ref{point32} has all momenta distinct 
form each other and exhibits the numerator 
$(\ell\cdot p_4)\times(\ell\cdot p_5)\times(\ell\cdot p_6)$.

\section{Conclusions}
\label{sec:conclusions}

The Loop-Tree Duality has many appealing theoretical properties for the calculation of processes with many external legs. In this paper, we have investigated the practicability of a first numerical implementation of the LTD method.

In our analysis of the singular behaviour of the loop integrand, we found two distinct types of singularities: Ellipsoid singularities which require the application of contour deformation and hyperboloid singularities that occur pairwise and cancel among different dual contributions. In order to preserve their cancellation, dual contributions featuring the same hyperboloid singularity pair must receive the same contour deformation. This leads to the following algorithm: Sets of pairwise coupled dual contributions are organized into groups. Each group is deformed independently from the others and each dual contribution of such a group receives the exact same contour deformation that accounts for all the ellipsoid singularities of the entire group. 

We applied a contour deformation that efficiently deals with the ellipsoid singularities by meeting all the important criteria \cite{Soper:1999xk}. This setup has been successful in the calculation of finite multi-leg scalar and tensor integrals. 
We found the results to be in very good agreement with the reference values produced by LoopTools and SecDec. An important further check of our implementation presented here was
various scans which show that the code handles equally well 
broad slices of the phase space. 
The code excels in cases that involve many external legs as it shows a modest increase in running times in comparison to cases with fewer legs. 
From this first study, we can be optimistic that our implementation of the 
LTD method offers a competitive
alternative for computing multi-scale, multi-leg scalar and tensor one-loop integrals.

In this paper we have only considered IR- and UV-finite integrals. However,
the treatment of IR- and UV-divergent graphs in the context of the LTD method 
and the description of how to combine directly the virtual with the 
real radiative corrections in order to obtain an infrared finite 
implementation has been presented recently~\cite{Hernandez-Pinto:2015ysa}.
The extension of our code to deal with these cases is an ongoing project.

\section*{Acknowledgements}

We thank S. Catani for a longstanding fruitful collaboration. We also thank 
S. Borowka and G. Heinrich for their help in running SecDec. 
This work has been supported by the Research Executive Agency (REA) 
under the Grant Agreement No. PITN-GA-2010-264564 (LHCPhenoNet), 
by the Spanish Government and ERDF funds from the European Commission (Grants
No. FPA2014-53631-C2-1-P, FPA2011-23778, FPA2013-44773-P),
and by Generalitat Valenciana under Grant No. PROMETEOII/2013/007.
G.C. acknowledges support from Marie Curie Actions (PIEF- GA-2011-298582).
S.B. acknowledges support from JAEPre programme (CSIC). P.D. acknowledges support from 
General Secretariat for Research and Technology of Greece and from European Regional Development Fund 
MIS-448332-ORASY (NSRF 2007-13 ACTION, KRIPIS).

\appendix

\section{External momenta and internal masses of the sample points} 
\label{AppendixA}

Here we give the external momenta and internal masses of the different phase-space points and 
scans shown in Sections \ref{sec:rs} and \ref{sec:rt}. 
Due to momentum conservation $p_N=-\sum_{i=1}^{N-1}p_i$, 
it is sufficient to give only the momenta $p_1$ to $p_{N-1}$.
Momenta and masses are implicitely given in GeV.\\
To produce the energy scans (Fig. \ref{fig:pentachu} and Fig. \ref{fig:pentanum}), we varied the 
external momentum $p_1$ by multiplying with the square root of some scaling parameter 
$\lambda$ (not to be confused with the scaling parameter $\lambda_{ij}$ of the contour 
deformation). This is indicated where we give the respective momenta.

\subsection{Individual sample points}

\begin{enumerate}[leftmargin=2.5cm]

  \item[{\bf Figure \ref{fig:trilms}}] 
$p_1=(44.38942,17.84418,12.70440,-23.67441)$\\
$p_2=(11.62982,-35.11756,-9.52573,1.27635)$\\
$m_1= m_2 = m_3 = 7.89824$

  \item[{\bf Figure \ref{fig:boxlms}}] 
$p_1=(95.95213, 65.25140,-40.62468,30.93648)$\\
$p_2=(68.47023,-60.09584,18.23998,84.29507)$\\
$p_3=(12.99839,12.08603,-99.08246,-34.58997)$\\
$m_1= m_2 = m_3 = m_4 = 11.50163$

  \item[{\bf P\ref{point1}}] 
$p_1=(5.23923,-4.18858,0.74966,-3.05669)$\\
$p_2=(6.99881,-2.93659,5.03338,3.87619)$\\
$m_1=m_2=m_3=7.73358$

  \item[{\bf P\ref{point2}}] 
$p_1=(13.42254,58.79478,-73.11858,-91.95015)$\\
$p_2=(81.65928,-68.52173,8.75578,-95.05353)$\\
$m_1= 49.97454, m_2=86.92490, m_3=80.22567$

  \item[{\bf P\ref{point3}}] 
$p_1=(10.51284,6.89159,-7.40660,-2.85795)$\\
$p_2=(6.45709,2.46635,5.84093,1.22257)$\\
$m_1= m_2 = m_3 = 0.52559$

  \item[{\bf P\ref{point4}}] 
$p_1=(95.77004,31.32025,-34.08106,-9.38565)$\\
$p_2=(94.54738,-53.84229,67.11107,45.56763)$\\
$m_1= 83.02643, m_2=76.12873, m_3=55.00359$

  \item[{\bf P\ref{point5}}] 
$p_1=(31.54872,-322.40325,300.53015,-385.58013)$\\
$p_2=(103.90430,202.00974,-451.27794,-435.12848)$\\
$p_3=(294.76653,252.88958,447.09194,311.71630)$\\
$m_1= m_2 = m_3 = m_4 = 4.68481$

  \item[{\bf P\ref{point6}}] 
$p_1=(50.85428,-55.74613,11.69987,94.92591)$\\
$p_2=(0.69914,67.19262,-5.78627,91.52776)$\\
$p_3=(52.35768,76.32258,43.82222,13.05874)$\\
$m_1= 54.29650, m_2 = 53.54058, m_3 = 55.96814, m_4 = 51.74438$

  \item[{\bf P\ref{point7}}] 
$p_1=(62.80274,-49.71968,-5.53340,-79.44048)$\\
$p_2=(48.59375,-1.65847,34.91140,71.89564)$\\
$p_3=(76.75934,-19.14334,-17.10279,30.22959)$\\
$m_1= m_2 = m_3 = m_4 = 9.82998$

  \item[{\bf P\ref{point8}}] 
$p_1=(98.04093, 77.37405, 30.53434,-81.88155)$\\
$p_2=(73.67657,-53.78754,13.69987,14.20439)$\\
$p_3=(68.14197,-36.48119,59.89499,-81.79030)$\\
$m_1= 81.44869, m_2 = 94.39003, m_3 = 57.53145, m_4 = 0.40190$

  \item[{\bf P\ref{point9}}] 
$p_1=(76.50219, -72.36197, 10.95225,-99.79612)$\\
$p_2=(99.02723,27.27133,-25.11907,86.10825)$\\
$p_3=(64.19420,13.10011,18.37737,-29.16095)$\\
$m_1= m_2 = 37.77809, m_3 = m_4 = 36.84323$

  \item[{\bf P\ref{point10}}] 
$p_1=(13.62303, -64.20757,-17.59085,-8.81785)$\\
$p_2=(96.67650,89.65623,-18.47276,40.73203)$\\
$p_3=(66.21913,-39.49917,3.640139,-82.31669)$\\
$m_1= m_3 = 64.67282, m_2 = m_4 = 51.13181$

  \item[{\bf P\ref{point11}}] 
$p_1=(33.74515, 45.72730,31.15254,-7.47943)$\\
$p_2=(31.36435,-41.50734,46.47897,2.04203)$\\
$p_3=(4.59005,17.07010,32.65403,41.93628)$\\
$p_4=(29.51054,-28.25963,46.17333,-35.08918)$\\
$m_1= m_2 = m_3 = m_4 = m_5 = 5.01213$

  \item[{\bf P\ref{point12}}] 
$p_1=(33.76482, 45.44063,-10.68084,16.41925)$\\
$p_2=(72.93498,67.49170,-11.81485,-36.28455)$\\
$p_3=(8.01673,-49.40112,-66.09200,-0.11414)$\\
$p_4=(-86.54188,-97.01228,68.12494,32.94875)$\\
$m_1= 98.42704, m_2 = 28.89059, m_3 = 40.51436, 
m_4 = 75.45643, m_5 = 11.08327$

  \item[{\bf P\ref{point13}}] 
$p_1=(1.58374, 6.86200,-15.06805,-10.63574)$\\
$p_2=(7.54800,-3.36539,34.57385,27.52676)$\\
$p_3=(43.36396,-49.27646,-25.35062,-17.68709)$\\
$p_4=(22.58103,38.31530,-14.67581,-3.08209)$\\
$m_1= m_2 = m_3 = m_4 = m_5 = 2.76340$

  \item[{\bf P\ref{point14}}] 
$p_1=(-89.85270,69.44839,-96.30496,14.47549)$\\
$p_2=(-81.61779,6.89065,1.76775,18.39834)$\\
$p_3=(-89.80789,24.32486,48.73341,0.74094)$\\
$p_4=(-43.20198,-85.34635,92.38148,93.84802)$\\
$m_1= 22.21430, m_2 = 15.84324, m_3 = 34.80431, 
m_4 = 27.53390, m_5 = 29.19823$

  \item[{\bf P\ref{point15}}] 
$p_1=(94.79774,-70.04005,-84.77221,36.09812)$\\
$p_2=(-42.15872,-36.33754,-14.72331,-41.24018)$\\
$p_3=(73.77293,88.37064,33.47296,-24.17542)$\\
$p_4=(81.85638,77.17370,-62.39774,-6.89737)$\\
$m_1= m_2 = m_3 = 1.30619, m_4 = m_5 = 1.26692$

  \item[{\bf P\ref{point24}}] 
$p_1=(69.70234,62.68042,25.44429,-97.78603)$\\
$p_2=(-65.98494,-85.19920,98.05702,-70.89141)$\\
$p_3=(-26.75642,-30.42288,-26.84633,14.81944)$\\
$p_4=(-69.44800,56.74842,-32.23649,96.45829)$\\
$m_1 = m_2 = m_3 = m_4 = m_5 = 87.00572$

  \item[{\bf P\ref{point25}}] 
$p_1=(-45.80756,95.63842,-55.04954,44.01174)$\\
$p_2=(36.09562,52.66752,-11.22354,-87.48918)$\\
$p_3=(-4.90798,41.11273,14.29379,2.15944)$\\
$p_4=(49.48233,40.26756,-23.16581,-96.89362)$\\
$m_1 = m_2 = m_3 = m_4 = m_5 = 56.97318$

  \item[{\bf P\ref{point26}}] 
$p_1=(-18.90057,-97.14671,44.69176,-16.67528)$\\
$p_2=(-70.86315,-81.27489,-3.71628,18.79403)$\\
$p_3=(-89.53092,50.02356,33.39784,-51.66031)$\\
$p_4=(-96.59097,-34.80215,-83.24353,44.73888)$\\
$m_1 = m_2 = m_3 = m_4 = m_5 = 43.87459$

  \item[{\bf P\ref{point27}}] 
$p_1=(-88.70322,37.98826,62.19352,-35.86433)$\\
$p_2=(-58.60617,-58.60074,-83.75298,61.78210)$\\
$p_3=(-83.73607,46.98912,67.44602,78.40612)$\\
$p_4=(-96.41508,71.69925,-14.47818,-61.82390)$\\
$m_1 = m_2 = m_3 = m_4 = m_5 = 16.73899$

  \item[{\bf P\ref{point28}}] 
$p_1=(-3.43584,4.73492,17.31242,61.53467)$\\
$p_2=(12.12233,32.23256,87.57836,-58.25073)$\\
$p_3=(-38.67209,-54.27020,21.15570,79.15640)$\\
$p_4=(-90.90573,-79.70266,-88.26463,-66.00973)$\\
$p_5=(-34.40043,-88.73043,84.41781,-4.21221)$\\
$m_1 = 54.36459, m_2 = 30.96600, m_3 = 51.03652,$\\
$m_4 = 16.03115, m_5 = 2.25657, m_6 = 59.45020$

  \item[{\bf P\ref{point29}}] 
$p_1=(-9.85384,15.70678,80.94234,-84.96387)$\\
$p_2=(90.11707,-74.59469,-70.73997,54.32748)$\\
$p_3=(-55.84212,-34.47531,-87.20597,-27.73882)$\\
$p_4=(16.72808,64.83574,-31.16733,63.94189)$\\
$p_5=(-42.62943,49.91058,-46.12974,59.76096)$\\
$m_1 = 42.61768, m_2 = 22.13590, m_3 = 34.87263,$\\
$m_4 = 54.00634, m_5 = 79.54844, m_6 = 87.50131$

  \item[{\bf P\ref{point30}}] 
$p_1=(35.27512,36.08798,-89.66662,18.22907)$\\
$p_2=(-32.58939,14.45447,86.93898,-47.20827)$\\
$p_3=(-76.40210,-62.22587,-63.59955,41.03465)$\\
$p_4=(-2.30248,0.45058,-76.74256,-64.19292)$\\
$p_5=(-88.80252,18.06504,-6.53891,49.34535)$\\
$m_1 = m_2 = m_3 = m_4 = m_5 = m_6 = 82.87370$

  \item[{\bf P\ref{point31}}] 
$p_1=(-99.20747,-68.16217,95.24772,68.87644)$\\
$p_2=(-95.09224,78.51258,-82.38270,20.36899)$\\
$p_3=(-56.04092,22.93681,-72.82681,96.81954)$\\
$p_4=(78.53840,-86.40143,-82.49674,-57.42855)$\\
$p_5=(13.70265,77.87278,99.79126,8.31677)$\\
$m_1 = 63.23680, m_2 = 86.48449, m_3 = 44.51361,$\\
$m_4 = 79.73599, m_5 = 74.43246, m_6 = 70.11421$

  \item[{\bf P\ref{point32}}] 
$p_1=(-70.26380,96.72681,21.66556,-37.40054)$\\
$p_2=(-13.45985,2.12040,3.20198,91.44246)$\\
$p_3=(-62.59164,-29.93690,-22.16595,-58.38466)$\\
$p_4=(-67.60797,-83.23480,18.49429,8.94427)$\\
$p_5=(-34.70936,-62.59326,-60.71318,2.77450)$\\
$m_1 = 94.53242, m_2 = 64.45092, m_3 = 74.74299,$\\
$m_4 = 10.63129, m_5 = 31.77881, m_6 = 23.93819$
\end{enumerate}

\subsection{Energy and mass scans}

\begin{enumerate}[resume*]
  \item[{\bf Figure \ref{fig:trithreshold}}] 
$p_1=(27.95884,25.55639,-29.88288,-2.17433)$\\
$p_2=(27.45521,-7.81292,3.19651,6.05088)$\\
$6.05088 \leq m_1 = m_2 = m_3 \leq 31.53414$

  \item[{\bf Figure \ref{fig:boxalleq}}] 
$p_1=(67.40483,49.44993,-20.67085,48.63654)$\\
$p_2=(54.64295,-58.23071,9.55042,-16.59411)$\\
$p_3=(41.37620,11.75178,-40.77655,-8.25014)$\\
$2.33822 \leq m_1 = m_2 = m_3 = m_4 \leq 70.14658$

  \item[{\bf Figure \ref{fig:pentachu}}] 
$p_{1}=\sqrt{\lambda} \, (-15.22437,-26.74156,6.65483,29.13661)~,
\qquad \lambda \in [1,30]~,$\\
$p_2=(-91.22611,-63.97875,55.07507,-52.90153)$\\
$p_3=(0.95105,75.90791,-10.13814,-88.40860)$\\
$p_4=(43.04908,77.11321,-50.69469,-7.60198)$\\
$m_1 = 49.12560, m_2 = 57.87487, m_3 = 26.47098, m_4 = 0.42094, m_5 = 62.31320$

  \item[{\bf Figure \ref{fig:pentanum}}] 
$p_{1}=\sqrt{\lambda} \, (-51.76504,-81.75539,-46.42422,-40.15540)~,
\qquad \lambda \in [1,30]~,$ \\
$p_{2}=(-63.76533,-2.53015,16.27485,69.16770)$\\
$p_{3}=(-78.50262,46.32052,13.19246,-54.00166)$\\
$p_{4}=(25.40582,81.48058,39.11105,93.24648)$\\
$m_1 = 78.45208, m_2 = 42.71315, m_3 = 91.94256, m_4 = 61.59730, m_5 = 16.75672$
\end{enumerate}

%%%%%%%%%%%%%%%%%%%%%%%%%%%%%%%%%%%%%%%%%%%%%%%%%%%%%%%%%%%%%%%%%%%
%%%%%%%%%%%%%%%%%%%%%%%%%%%%%%%%%%%%%%%%%%%%%%%%%%%%%%%%%%%%%%%%%%%
%%%%%%%%%%%%%%%%%%%%%%%%%%%%%%%%%%%%%%%%%%%%%%%%%%%%%%%%%%%%%%%%%%%


\begin{thebibliography}{90}
  
%%%%% Numerical %%%%

\bibitem{Catani:1996jh} 
  S.~Catani and M.~H.~Seymour,
  ``The Dipole formalism for the calculation of QCD jet cross-sections at next-to-leading order,''
  Phys.\ Lett.\ B {\bf 378}, 287 (1996)
  [hep-ph/9602277].
  
\bibitem{Soper:1998ye}
  D.~E.~Soper,
  ``QCD calculations by numerical integration,''
  Phys.\ Rev.\ Lett.\  {\bf 81} (1998) 2638
  [hep-ph/9804454].
  %%CITATION = HEP-PH/9804454;%%

\bibitem{Soper:1999xk}
  D.~E.~Soper,
  ``Techniques for QCD calculations by numerical integration,''
  Phys.\ Rev.\ D {\bf 62} (2000) 014009
  [hep-ph/9910292].
  %%CITATION = HEP-PH/9910292;%%

\bibitem{Soper:2001hu}
  D.~E.~Soper,
  ``Choosing integration points for QCD calculations by numerical integration,''
  Phys.\ Rev.\ D {\bf 64} (2001) 034018
  [hep-ph/0103262].
  %%CITATION = HEP-PH/0103262;%%

\bibitem{Kramer:2002cd}
  M.~Kramer and D.~E.~Soper,
  ``Next-to-leading order numerical calculations in Coulomb gauge,''
  Phys.\ Rev.\ D {\bf 66} (2002) 054017
  [hep-ph/0204113].
  %%CITATION = HEP-PH/0204113;%%

\bibitem{Ferroglia:2002mz}
  A.~Ferroglia, M.~Passera, G.~Passarino and S.~Uccirati,
  ``All purpose numerical evaluation of one loop multileg Feynman diagrams,''
  Nucl.\ Phys.\ B {\bf 650} (2003) 162
  [hep-ph/0209219].
  %%CITATION = HEP-PH/0209219;%%

\bibitem{Nagy:2003qn}
  Z.~Nagy and D.~E.~Soper,
  ``General subtraction method for numerical calculation of one loop QCD matrix elements,''
  JHEP {\bf 0309} (2003) 055
  [hep-ph/0308127].
  %%CITATION = HEP-PH/0308127;%%

\bibitem{Nagy:2006xy}
  Z.~Nagy and D.~E.~Soper,
  ``Numerical integration of one-loop Feynman diagrams for N-photon amplitudes,''
  Phys.\ Rev.\ D {\bf 74} (2006) 093006
  [hep-ph/0610028].
  %%CITATION = HEP-PH/0610028;%%

\bibitem{Moretti:2008jj}
  M.~Moretti, F.~Piccinini and A.~D.~Polosa,
  ``A Fully Numerical Approach to One-Loop Amplitudes,''
  arXiv:0802.4171 [hep-ph].
  %%CITATION = ARXIV:0802.4171;%%

\bibitem{Gong:2008ww}
  W.~Gong, Z.~Nagy and D.~E.~Soper,
  ``Direct numerical integration of one-loop Feynman diagrams for N-photon amplitudes,''
  Phys.\ Rev.\ D {\bf 79} (2009) 033005
  [arXiv:0812.3686 [hep-ph]].
  %%CITATION = ARXIV:0812.3686;%%

\bibitem{Kilian:2009wy}
  W.~Kilian and T.~Kleinschmidt,
  ``Numerical Evaluation of Feynman Loop Integrals by Reduction to Tree Graphs,''
  arXiv:0912.3495 [hep-ph].
  %%CITATION = ARXIV:0912.3495;%%

\bibitem{Becker:2010ng}
  S.~Becker, C.~Reuschle and S.~Weinzierl,
  ``Numerical NLO QCD calculations,''
  JHEP {\bf 1012} (2010) 013
  [arXiv:1010.4187 [hep-ph]].
  %%CITATION = ARXIV:1010.4187;%%

\bibitem{Becker:2012aqa}
  S.~Becker, C.~Reuschle and S.~Weinzierl,
  ``Efficiency Improvements for the Numerical Computation of NLO Corrections,''
  JHEP {\bf 1207} (2012) 090
  [arXiv:1205.2096 [hep-ph]].
  %%CITATION = ARXIV:1205.2096;%%

\bibitem{Becker:2012nk}
  S.~Becker and S.~Weinzierl,
  ``Direct contour deformation with arbitrary masses in the loop,''
  Phys.\ Rev.\ D {\bf 86} (2012) 074009
  [arXiv:1208.4088 [hep-ph]].
  %%CITATION = ARXIV:1208.4088;%%  

\bibitem{Bevilacqua:2011xh} 
  G.~Bevilacqua, M.~Czakon, M.~V.~Garzelli, A.~van Hameren, A.~Kardos, C.~G.~Papadopoulos, R.~Pittau and M.~Worek,
  ``Helac-nlo,''
  Comput.\ Phys.\ Commun.\  {\bf 184}, 986 (2013)
  [arXiv:1110.1499 [hep-ph]].
  
\bibitem{Cascioli:2011va}
  F.~Cascioli, P.~Maierhofer and S.~Pozzorini,
  ``Scattering Amplitudes with Open Loops,''
  Phys.\ Rev.\ Lett.\  {\bf 108} (2012) 111601
  [arXiv:1111.5206 [hep-ph]].
  %%CITATION = ARXIV:1111.5206;%%

\bibitem{Cullen:2014yla} 
  G.~Cullen {\it et al.},
  ``G$\scriptsize{O}$S$\scriptsize{AM}$-2.0: a tool for automated one-loop calculations within the Standard Model and beyond,''
  Eur.\ Phys.\ J.\ C {\bf 74}, no. 8, 3001 (2014)
  [arXiv:1404.7096 [hep-ph]].
  
\bibitem{Gleisberg:2008ta} 
  T.~Gleisberg, S.~Hoeche, F.~Krauss, M.~Schonherr, S.~Schumann, F.~Siegert and J.~Winter,
  ``Event generation with SHERPA 1.1,''
  JHEP {\bf 0902}, 007 (2009)
  [arXiv:0811.4622 [hep-ph]].

\bibitem{Frixione:2008ym} 
  S.~Frixione and B.~R.~Webber,
  ``The MC and NLO 3.4 Event Generator,''
  arXiv:0812.0770 [hep-ph].
  
\bibitem{Alwall:2014hca} 
  J.~Alwall {\it et al.},
  ``The automated computation of tree-level and next-to-leading order differential cross sections, and their matching to parton shower simulations,''
  JHEP {\bf 1407}, 079 (2014)
  [arXiv:1405.0301 [hep-ph]].  

 
%%%%%%%%%%%%%%%%%%%%%%%%%%%%%%%%%%

\bibitem{Passarino:2001wv}
  G.~Passarino,
  ``An Approach toward the numerical evaluation of multiloop Feynman diagrams,''
  Nucl.\ Phys.\ B {\bf 619} (2001) 257
  [hep-ph/0108252].
  %%CITATION = HEP-PH/0108252;%%

\bibitem{Anastasiou:2007qb}
  C.~Anastasiou, S.~Beerli and A.~Daleo,
  ``Evaluating multi-loop Feynman diagrams with infrared and threshold singularities numerically,''
  JHEP {\bf 0705} (2007) 071
  [hep-ph/0703282].
  %%CITATION = HEP-PH/0703282;%%

\bibitem{Becker:2012bi}
  S.~Becker and S.~Weinzierl,
  ``Direct numerical integration for multi-loop integrals,''
  Eur.\ Phys.\ J.\ C {\bf 73} (2013) 2321
  [arXiv:1211.0509 [hep-ph]].
  %%CITATION = ARXIV:1211.0509;%%

%%%% loop--tree duality %%%%

\bibitem{Catani:2008xa}
  S.~Catani, T.~Gleisberg, F.~Krauss, G.~Rodrigo and J.~C.~Winter,
  ``From loops to trees by-passing Feynman's theorem,''
  JHEP {\bf 0809} (2008) 065
  [arXiv:0804.3170 [hep-ph]].
  %%CITATION = ARXIV:0804.3170;%%

\bibitem{Rodrigo:2008fp}
  G.~Rodrigo, S.~Catani, T.~Gleisberg, F.~Krauss and J.~C.~Winter,
  ``From multileg loops to trees (by-passing Feynman's Tree Theorem),''
  Nucl.\ Phys.\ Proc.\ Suppl.\  {\bf 183} (2008) 262
  [arXiv:0807.0531 [hep-th]].
  %%CITATION = ARXIV:0807.0531;%%

\bibitem{Bierenbaum:2010cy}
  I.~Bierenbaum, S.~Catani, P.~Draggiotis and G.~Rodrigo,
  ``A Tree-Loop Duality Relation at Two Loops and Beyond,''
  JHEP {\bf 1010} (2010) 073
  [arXiv:1007.0194 [hep-ph]].
  %%CITATION = ARXIV:1007.0194;%%

\bibitem{Bierenbaum:2012th}
  I.~Bierenbaum, S.~Buchta, P.~Draggiotis, I.~Malamos and G.~Rodrigo,
  ``Tree-Loop Duality Relation beyond simple poles,''
  JHEP {\bf 1303} (2013) 025
  [arXiv:1211.5048 [hep-ph]].
  %%CITATION = ARXIV:1211.5048;%%

\bibitem{Bierenbaum:2013nja}
  I.~Bierenbaum, P.~Draggiotis, S.~Buchta, G.~Chachamis, I.~Malamos and G.~Rodrigo,
  ``News on the loop--tree Duality,''
  Acta Phys.\ Polon.\ B {\bf 44} (2013) 2207.
  %%CITATION = APPOA,B44,2207;%%
  
\bibitem{Buchta:2014dfa}
  S.~Buchta, G.~Chachamis, P.~Draggiotis, I.~Malamos and G.~Rodrigo,
  ``On the singular behaviour of scattering amplitudes in quantum field theory,''
  JHEP {\bf 1411} (2014) 014
  [arXiv:1405.7850 [hep-ph]].
  %%CITATION = ARXIV:1405.7850;%%

\bibitem{Buchta:2014fva}
  S.~Buchta, G.~Chachamis, I.~Malamos, I.~Bierenbaum, P.~Draggiotis and G.~Rodrigo,
  ``The loop-tree duality at work,''
  PoS LL {\bf 2014} (2014) 066
  [arXiv:1407.5865 [hep-ph]].
  %%CITATION = ARXIV:1407.5865;%%

\bibitem{Buchta:phd}
  S.~Buchta, PhD thesis, Universitat de Val\`encia, 2015,
  %S.~Buchta,
  ``Theoretical foundations and applications of the Loop-Tree Duality in Quantum Field Theories,''
  [arXiv:1509.07167 [hep-ph]].
  %%CITATION = ARXIV:1509.07167;%%

\bibitem{Buchta:2015vha}
  S.~Buchta, G.~Chachamis, P.~Draggiotis, I.~Malamos and G.~Rodrigo,
  ``Towards a Numerical Implementation of the Loop-Tree Duality Method,''
  Nucl.\ Part.\ Phys.\ Proc.\  {\bf 258-259} (2015) 33
  [arXiv:1509.07386 [hep-ph]].
  %%CITATION = 10.1016/j.nuclphysbps.2015.01.008;%%

\bibitem{Hernandez-Pinto:2015ysa}
  R.~J.~Hern\'andez-Pinto, G.~F.~R.~Sborlini and G.~Rodrigo,
  ``Gauge theories in four dimensions,''
  arXiv:1506.04617 [hep-ph].
  %%CITATION = ARXIV:1506.04617;%%

%%%%%%%%%%%%%%%%%%%%%%%%%%%%%%%%%%%%%%%%%%%%%%%%%%%%

%\cite{Hahn:1998yk}
\bibitem{Hahn:1998yk} 
  T.~Hahn and M.~Perez-Victoria,
  %``Automatized one loop calculations in four-dimensions and D-dimensions,''
  Comput.\ Phys.\ Commun.\  {\bf 118}, 153 (1999)
  [hep-ph/9807565].
  
\bibitem{Hahn:2004fe}
  T.~Hahn,
  ``CUBA: A Library for multidimensional numerical integration,''
  Comput.\ Phys.\ Commun.\  {\bf 168} (2005) 78
  [hep-ph/0404043].
  %%CITATION = HEP-PH/0404043;%%   
  
\bibitem{Cuhre1}
  J.~Berntsen, T.~O.~Espelid, A.~Genz,
  ''An Adaptive Algorithm for the Approximate Calculation of Multiple Integrals''
  ACM Trans. Math. Softw. {\bf 17} (1991) 437-451.
  
\bibitem{Cuhre2}
  J.~Berntsen, T.~O.~Espelid, A.~Genz,
  ''An Adaptive Multidimensional Integration Routine for a Vector of Integrals''
  ACM Trans. Math. Softw. {\bf 17} (1991) 452-456.  

\bibitem{Lepage:1980dq}
  G.~P.~Lepage,
  ``Vegas: An Adaptive Multidimensional Integration Program,''
  Report No CLNS-80/447.
  %%CITATION = CLNS-80/447;%%
  
\bibitem{Mathematica}
  Wolfram Research, Inc.,
  ``Mathematica,''
   Version 10.0, (2015).
  
\bibitem{Borowka:2015mxa}
  S.~Borowka, G.~Heinrich, S.~P.~Jones, M.~Kerner, J.~Schlenk and T.~Zirke,
  ``SecDec-3.0: numerical evaluation of multi-scale integrals beyond one loop,''
  arXiv:1502.06595 [hep-ph].
  %%CITATION = ARXIV:1502.06595;%%

\end{thebibliography}
\end{document}